\documentclass[12pt, draftclsnofoot, onecolumn]{IEEEtran}
\usepackage{amsmath,amssymb,epsfig,amsthm,psfrag,epsf, multirow,wrapfig, graphicx}
\usepackage[english]{babel}
\usepackage{epstopdf}
\usepackage[utf8]{inputenc}
\usepackage{cite}

\usepackage{cleveref}
\usepackage{ dsfont }
\usepackage{changes}
\usepackage{bm}
\usepackage{algorithm}
\usepackage[]{algorithmicx}
\usepackage{algpseudocode} 
\usepackage{tikz}
\usetikzlibrary{shapes.geometric, arrows}
\tikzstyle{startstop} = [rectangle, rounded corners, minimum width=3cm, minimum height=1cm,text centered, draw=black, fill=red!30]
\tikzstyle{io} = [trapezium, trapezium left angle=70, trapezium right angle=110, minimum width=3cm, minimum height=1cm, text centered, draw=black, fill=blue!30]
\tikzstyle{process} = [rectangle, minimum width=3cm, minimum height=1cm, text centered, draw=black, text width=12cm, fill=blue!20]
\tikzstyle{decision} = [diamond, minimum width=3cm, minimum height=1cm, text centered, draw=black, text width=2cm, fill=green!30]
\tikzstyle{arrow} = [thick,->,>=stealth]

\algnewcommand\algorithmicinitialization{\textbf{STEP 1: Initialization}}
\algnewcommand\INITIALIZATION{\item[\algorithmicinitialization]}
\algnewcommand\algorithmicoptimization{\textbf{STEP 2: Optimization of Beam Weights}}
\algnewcommand\OPTIMIZATION{\item[\algorithmicoptimization]}
\algnewcommand\algorithmiclearn{\textbf{STEP 2: Learning Best Beams}}
\algnewcommand\LEARN{\item[\algorithmiclearn]
}

\usepackage{caption}
\usepackage{subcaption}
\usepackage{color}

\title{ 
Self-Tuning Sectorization: Deep Reinforcement Learning Meets Broadcast Beam Optimization}
\author{
\IEEEauthorblockN{Rubayet Shafin, Hao Chen, Young Han Nam, Sooyoung Hur, Jeongho Park, Jianzhong (Charlie) Zhang, Jeffrey Reed, and Lingjia Liu}
\thanks{R. Shafin, J. Reed, and L. Liu are with Wireless@VT, Bradley Department of Electrical and Computer Engineering, Virginia Tech, Blacksburg, VA 24061, USA; H. Chen, Y. H. Nam, and J.  Zhang are with the Standards and 5G Mobility Laboratory of Samsung Research America, Richardson, Texas, 75082, USA; S. Hur and J. Park are with Air Technology Lab, Samsung Electronics, Korea. This work was done in part while the first author was with Samsung Research America.}
\thanks{The corresponding author is L. Liu (ljliu@ieee.org). This work has been submitted to the IEEE for possible publication. Copyright may be transferred without notice, after which this version may no longer be accessible.}
}
\begin{document}
	\maketitle
% \thispagestyle{plain}
% \pagestyle{plain}	
	%\thispagestyle{firstpage}
	%\newpage
	
\begin{abstract}
Beamforming in multiple input multiple output (MIMO) systems is one of the key technologies for modern wireless communication. Creating appropriate sector-specific broadcast beams are essential for enhancing the coverage of cellular network and for improving the broadcast operation for control signals. However, in order to maximize the coverage, patterns for broadcast beams need to be adapted based on the users' distribution and movement over time. In this work, we present self-tuning sectorization: a deep reinforcement learning framework to optimize MIMO broadcast beams autonomously and dynamically based on user' distribution in the network. Taking directly UE measurement results as input, deep reinforcement learning agent can track and predict the UE distribution pattern and come up with the best broadcast beams for each cell. Extensive simulation results show that the introduced framework can achieve the optimal coverage, and converge to the oracle solution for both single sector and multiple sectors environment, and for both periodic and Markov mobility patterns.
\end{abstract}

	
\section{Introduction}
Cellular data traffic has witnessed an exponential growth over the last few years primarily due to the widespread use of mobile devices and novel application services. Cisco Visual Networking Index (VNI) forecast predicts a threefold increase of global IP traffic from 122 exabyte (EB) in 2017 to 296 EB in 2022\cite{Cisco_VNI2018}. In order to handle this massive data-flow and ensure superior quality of experience (QoE) to the end users, wireless cellular networks are also becoming extremely complicated. With the coexistence of different types of networks, managing networks efficiently has become a critical issue for 5G~\cite{andrews2014will,boccardi2014five,wang2014cellular} and beyond systems. In order to reduce the network management complexity and operational cost, self organizing network (SON) has been introduced in Third Generation Partnership Project (3GPP) as one of the enabling technologies for advanced mobile networks~\cite{hamalainen2012lte, peng2013self}. SON aims to achieve autonomous functionalities within Radio Access Network (RAN). These self-X functionalities include self-configuration, self-optimization, and self-healing~\cite{sallent2011roadmap,hu2010self}. Self-optimization within SON refers to the process of self-tuning of network parameters for achieving optimum performance in terms of any predefined metric of interest. The idea is to dynamically update the cellular radio resource parameters based on the changes in propagation characteristics, traffic pattern or network deployment scenarios.
User distribution in wireless cellular network changes dynamically over time. These changes are the result of users' mobility behavior. For instance, in the day time,  users are more densely populated in the commercial area whereas at night, users are primarily clustered in residential areas. Users' large time-scale movement also depends on specific time within the week (workdays and weekends) or year (holidays). Accordingly, to maximize the overall throughput and coverage of the wireless networks, sector-specific cellular radio parameters should also be updated taking into account the changes in users' distribution. 
	
Multiple Input Multiple Output (MIMO) system~\cite{Downlink_MIMO} is one of the back-bones for current and next generation cellular network. Massive MIMO \cite{Noncooperative_Cellular}, where a large number of antennas are deployed at the base stations (BS), is envisioned as a key enabler for 5G systems. Beamforming refers to a MIMO technique for coherently combining the signals generated by multiple antennas in the MIMO arrays. 3-dimensional (3D) massive MIMO/full-dimension (FD) MIMO~\cite{Full_Dimension_MIMO, Rubayet_Journal, RubayetJournal3} promises tremendous throughput gain by enabling simultaneous beamforming in both elevation and azimuth domain. With large antenna array, it is possible to create sharp narrow beams towards desired users, and hence reduce the interference significantly~\cite{RubayetJournal2}; this beamforming is used to improve user's throughput and is therefore user-specific. 
Cellular networks, on the other hand, also require to create wide beams.
In fact, sectorization can be viewed as a process of expansive beam generation where a separate wide beam is used to cover a separate sector belonging to the same cell-site. 
These sector-specific broad beams are essential for connecting as many users as possible, 
essentially providing the coverage for cellular networks. Another important application for wide beams is the broadcast technologies for sending out the wireless control and access signals as prescribed by LTE and LTE-Advanced systems. 
As a result, generating the accurate wide broadcast beam patterns that cover the maximum number of users in the network is critical.
	
Unfortunately, most of the works in the MIMO literature focus on maximizing MIMO throughput or increasing the reliability of the data plane. Meanwhile, at present, broadcast beam parameters are set manually in modern cellular networks: a group of network engineers do the drive tests and physically visit each base station site to fix the parameters controlling the shape, tilt and beam-widths of these sector-specific broadcast beams. Once fixed, these broadcast beam parameters are not changed until some major fault/complain is reported. In other words, the parameters remain unchanged for a long period of time-- often years, and as a result, currently, these parameter cannot be updated based on users' movement or change in user distribution. Accordingly, this fixed parameter setup results in strictly suboptimal solution in terms of overall network coverage. 

Reinforcement learning (RL) is shown to be a useful tool for dynamic spectrum access (DSA) as well as small cell networks. A Q-learning based framework has been introduced in \cite{galindo2010distributed} for managing cumulative interference, originated from multiple cognitive radios, at the primary users' receivers in wireless regional area networks (WRANs). The introduced RL system is shown to autonomously learn policy that handles the cumulative interference at the primary users and keeps interference level at the primary protection contour below a predefined threshold. An RL-based power control strategy has been developed in \cite{saad2012distributed} for cognitive femtocell networks, and it has been shown that RL can enhance the capacity of femtocells while ensuring a minimum quality of service (QoS) to macrocells. In a similar setup, \cite{bennis2013self} proposes  an RL framework for interference management in small-cell networks. The problem of dynamic channel assignment (DCA) has been addressed in \cite{nie1999q} by utilizing a real-time RL-based approach. A Multirate transmission control (MTC) strategy has been proposed in \cite{chen2004q} using Q-learning algorithm for wideband code division multiple access (CDMA) systems.

In cellular networks, users movement changes in a dynamic fashion. 
To maximize the coverage area, the braodcast beam parameters controlling cell-sectorization need to be dynamically updated based on user movement. However, selecting the best broadcast beams simultaneously for all sectors in the network and updating the beams autonomously based on users' movement or distribution is a challenging problem, primarily because of the large number of combinatorial possibilities for beam selection--this pertains to very large action space in RL framework.
Recently deep reinforcement learning (DRL) \cite{sutton2018reinforcement, mnih2013playing} has been proved to be capable of learning human-level control policies on a varieties of different Atari games \cite{mnih2015human}. DRL agents learn to estimate the Q-values of selecting the best possible actions from current state of the video games. However, compared to traditional Q-learning, in deep learning based Q-network, the Q-values are approximated using deep neural network instead of storing the Q-values for all state-action pairs in a tabular form. 
As a result, DRL has the ability to predict the correct Q-values even for very large state and action space.
Our recent work~\cite{Hao-HsuanDSA_RC18} shows that DRL based resource allocation can help improve the network performance of a DSA network.

In this work, we present a DRL-based framework for MIMO broadcast beam optimization for optimal cell-sectorization in order to maximize the coverage of cellular network. This will be an important step towards realizing the potential of SON. 
Our detailed contributions in this paper are summarized below:
\begin{enumerate}
\item We propose a double deep Q-network (DQN)-based framework \cite{van2016deep} for dynamically optimizing sector-specific MIMO broadcast beams for cellular network. The proposed learning-based algorithm can autonomously update or self-tune the beam parameters based on users' mobility patterns or changes in user distribution. 
\item We introduce self-tuning sectorization algorithms  for both single sector and multiple sector environments. For multiple sector case, we have proposed a novel neural network architecture for computing the Q-values corresponding to different broadcast beam selection, with only linear increase in complexity as the number of BSs increases.
\item Finally, we present extensive simulation work using ray-tracing data for validating our proposed solution. We consider both periodic and Markov mobility patterns, and show that the proposed DRL-based algorithm can achieve perfect convergence with \textit{Oracle} for both single sector and multiple sector environment and for any user distribution.
\end{enumerate}
The rest of the paper is organized as follows: Section \ref{System_Model} presents the network model and problem statement; Section \ref{Learning_Framework} presents the beam learning framework; Section \ref{DRL_optimization} introduces the DRL-based optimization strategies for both single cell and multiple cell environments; Section \ref{Simulation_section} presents the simulation work before we conclude the paper in Section \ref{conclusion}.
	
\section{Network Model and Problem Statement}\label{System_Model}
%	\begin{figure}[h!]
%		\centering
%		\includegraphics[width=0.8\linewidth]{Figures/system}
%		%\captionsetup{margin= {20pt},justification=centerlast,skip=-5pt,font=normalsize}
%		%\vspace{0.1cm}
%		\caption[width=0.5\linewidth]{{Network Model and Problem Statement.}}
%		%\captionsetup{justification=centering}
%		\label{system}
%	\end{figure}
We consider a cellular network consisting of $G$ BSs and $K$ UEs. We assume the BSs can have one or multiple sectors, and there are total $M$ sectors in the network, where $M \geq G$. Each sector is equipped with a two dimensional (2D) antenna array whose phases can be configured so that different array-beam widths (in both elevation and azimuth domain) and elevation tilt (e-tilt) angle can be updated. Placing 2D antenna array enables the BSs to beamform in both elevation and azimuth directions, and this is essentially the setup for full dimension (FD) MIMO systems \cite{Full_Dimension_MIMO, nam2013full}. The elevation beam-width, $\phi$, azimuth beam-width, $\psi$, and e-tilt angle, $\zeta$, constitute the parameter set in constructing the broadcast beams for each sector.  
	In this work, we focus on optimizing the broadcast beams/sector-wide beams for cellular network. 
Let us denote the number of antenna elements in elevation and azimuth directions by $N_1$ and $N_2$, respectively. Hence, total $N=N_1N_2$ number of antenna weights need to be tuned for generating the FD-MIMO broadcast beams. We can represent the $N_1 \times N_2$ antenna weight matrix into a $N \times 1$ weight vector, $\mathbf{w}$, following a vectorization operation. Each choice of weight vector, $\mathbf{w}$, in fact, consists of a specific choice of $\phi$, $\psi$, and $\zeta$. 
% For illustration, a particular choice of the parameters is shown in phasor form in Fig.~\ref{Phasor}.
A collection of notations used in this paper is summarized in Table \ref{table:notation}.
\begin{table}[ht]
% \small
    \caption{Notation for System Variables}
    \centering
    \begin{tabular}{c c}
    \hline\hline
    Variable & Notation \\ [0.5ex] % inserts table %heading
    \hline
    No. of BSs & $G$ \\
    No. of Sectors & $M$ \\
    No. of UEs & $K$ \\
    Elevation beam-width & $\phi$\\
    Azimuth beam-width & $\psi$\\
    E-tilt angle & $\zeta$\\
    No. of antennas at the BSs in elevation direction & $N_1$\\
    No. of antennas at the BSs in azimuth direction & $N_2$\\
    Total no. of antennas & $N$\\
    Broadcast signal from $m$-th BS & $x_m$\\
    Broadcast beamforming vector for $m$-th BS & $\mathbf{f}_m$\\
    Received signal at $k$-th UE & $y_k$\\
    Channel between $m$-th BS and $k$-th UE & $\mathbf{h}_{m,k}$\\
    Beam-pool & $\mathcal{W}$\\
    No. of possible beams in beam-pool & $J$\\
    $j$-th beam-weight vector in beam-pool & $\mathbf{w}^j$\\
    $n$-th antenna weight in $j$-th beam & $w_n^j$\\
    UEs' SINR threshold for connectivity & $T$\\
    \hline
    \end{tabular}
    \label{table:notation}
\end{table}

Assuming each UE has a  single antenna, the downlink broadcast received signal at $k$-th UE under $m$-th cell-sector can be written as 
\begin{align}\label{y_mk}
y_{k}=\mathbf{h}_{m,k}^T\mathbf{f}_{m}x_m+\sum_{\substack{m'=1\\m'\neq m}}^{M}\mathbf{h}_{m',k}^T\mathbf{f}_{m'}x_{m'}+n_{k},
\end{align}
where $\mathbf{h}_{m,k}$ is the $N \times 1$ channel vector for the channel between $m$-th sector and the $k$-th UE, $x_m$ is the broadcast signal from $m$-th sector, and $\mathbf{f}_m$ is the corresponding $N \times 1$ broadcast precoding vector for $m$-th sector. It can be clearly observed from \eqref{y_mk} that broadcast beams from one sector interfere with the beams from other sectors. Hence, in order to maximize the network coverage, selecting the appropriate broadcast beams for all the sectors is critical.
% \begin{figure}[h!]
% \centering
% \includegraphics[width=0.8\linewidth]{Figures/Other/Phasor}
% %\captionsetup{margin= {20pt},justification=centerlast,skip=-5pt,font=normalsize}
% %\vspace{0.1cm}
% \caption[width=0.5\linewidth]{{Example choice for elevation and azimuth beam-widths, and elevation tilt angle.}}
% %\captionsetup{justification=centering}
% \label{Phasor}
% \end{figure}
	
In this work, we adopt a DRL-based approach where an agent is  responsible for selecting the proper antenna configurations for all sectors. Each BS, for its sectors,  has the same pool of possible antenna weight vectors available, $\mathcal{W}:\{\mathbf{w}^1, \mathbf{w}^2, \ldots, \mathbf{w}^{J}\}$, where $J$ is the total number of beam-weight vectors in the pool; $\mathbf{w}^j=[w^j_1, w^j_2, \ldots, w^j_{N}]$ is the $j$-th vector in the beam pool, and $w^q_n$ is the antenna weight for the $n$-th antenna element corresponding to $q$-th weight vector. Accordingly, each sector  chooses its precoder, $\mathbf{f}$,  from  the beam pool, i.e., $\mathbf{f}_m \in \mathcal{W}$. It is to be noted here again that each of the weight vector in the pool corresponds to a particular choice of elevation and azimuth beam-widths and e-tilt angle. The agent  selects one out of $J$ beam patterns for each sector based on users' distribution/mobility patterns. 
This selection behavior is referred to as \textit{actions} in reinforcement learning. 
% \begin{figure}[h!]
% \centering
% \includegraphics[width=1.0\linewidth]{Figures/pool}
% %\captionsetup{margin= {20pt},justification=centerlast,skip=-5pt,font=normalsize}
% %\vspace{0.1cm}
% \caption[width=0.5\linewidth]{{Antenna Beam Pool}}
% %\captionsetup{justification=centering}
% \label{pool}
% \end{figure}

All BS in the network  transmit sector-specific signals using the wide broadcast beams selected by the agent. UEs collect measurement results such as Reference Signal Received Power (RSRP) or Reference Signal Received Quality (RSRQ), and report them to the agent as observation of the mobile environment. Assuming $k$-th UE in the network is associated with $m$-th sector, from \eqref{y_mk}, the received signal-to-interference-plus-noise ratio (SINR) for $k$-th user can be expressed as:
\begin{align}
\text{SINR}_{k}=\frac{\left|\mathbf{h}_{m,k}^T\mathbf{f}_{m}\right|^2}{\sum_{\substack{m'=1\\m'\neq m}}^{M}\left|\mathbf{h}_{m',k}^T\mathbf{f}_{m'}\right|^2+\sigma^2},
\end{align}
where $\sigma^2$ is the noise variance. In this work, we use the number of connected UEs as a metric to measure the cell coverage.
Number of connected UEs in the network can be defined as the number of UEs whose received signal-to-interference-plus-noise ratio (SINR) are above a predefined threshold, $T$. For any user distribution, the objective, hence, is to select the optimal beam pattern indices for all the sectors under all BSs that maximize the coverage or total number of connected UEs in the network. The problem can formally be written as:
	
\begin{align}
\max_{\mathbf{f}_1, \mathbf{f}_2, \ldots, \mathbf{f}_M} \sum_{\substack{k=1}}^{K} \mathds{1}_{\text{SINR}_{k}>T}\\
\text{subject to  \ \ \ \ \ \ \ \  } \mathbf{f}_m \in \mathcal{W}, \ \ \ \ 1\leq m \leq M,
\end{align}
where the indicator function, $\mathds{1}_{x>T}$, is defined as
\begin{align}
\mathds{1}_{x>T}:=\begin{cases}
1, & \text{if  } x>T\\
0,  & \text{if  } x \leq T.
\end{cases}
\end{align}
The user distribution changes over time, and hence optimal beam patterns that maximize the number of connected UEs at time $t_1$ may not be the  same as that at time $t_2$, where $t_1 \neq t_2$. The agent, therefore, has to be  able to  identify users' mobility pattern, and then  dynamically and autonomously select optimal beams for all the sectors in order to maximize network coverage. It is to be noted here that we are not using  users' location information to optimize the beam patterns. In order to minimize the feedback from the network, the agent will be merely using users' RSRP values to for the optimization.

In this work, we consider both single cell and multiple cells network scenarios. In the single cell case, the agent optimizes the broadcast beam for one cell--this represents a noise-limited environment. In this case, the DRL only needs to learn the optimal beam according to the cell environment including UE mobility pattern. On the other hand, in the multiple-cell case, the broadcast beams for all the cells need to be updated simultaneously-- this represents an interference-limited environment. We are addressing the challenges of these two scenarios where UEs are assumed to be moving according to some mobility pattern; first, the periodic case, where users' movement change in a periodic fashion, and second, the Markov case, where users' mobility is determined by following a transition probability matrix.
	
%	 two application scenarios: static scenario where the UEs are assumed to be static; dynamic scenario where UEs are assumed to be moving according to some mobility pattern. Each UE is assumed to connect to the BS with the highest received power. A UE is assumed to be connected if the received signal-to-interference-plus-noise ratio (SINR) is above a threshold T.   Each BS with $N$ antennas has a weight-parameter vector $[w_1, w_2, \ldots, w_N]$ as shown in Fig.~\ref{antenna}. The same pool of $k$ such antenna weight vectors are available to all the BSs as shown in Fig.~\ref{pool}
%	\begin{figure}[h!]
%		\centering
%		\includegraphics[width=0.8\linewidth]{Figures/antenna}
%		%\captionsetup{margin= {20pt},justification=centerlast,skip=-5pt,font=normalsize}
%		%\vspace{0.1cm}
%		\caption[width=0.5\linewidth]{{Weight Parameters for BS Antenna Array}}
%		%\captionsetup{justification=centering}
%		\label{antenna}
%	\end{figure}
%	\begin{figure}[h!]
%		\centering
%		\includegraphics[width=1.0\linewidth]{Figures/pool}
%		%\captionsetup{margin= {20pt},justification=centerlast,skip=-5pt,font=normalsize}
%		%\vspace{0.1cm}
%		\caption[width=0.5\linewidth]{{Antenna Weight Pool}}
%		%\captionsetup{justification=centering}
%		\label{pool}
%	\end{figure}
\section{Learning Framework}\label{Learning_Framework}
In this section, we present the learning framework for MIMO broadcast beam optimization using DRL as a self-tuning sectorization mechanism. We first briefly describe the background of DRL which will set up the foundation for the proposed broadcast beam-learning strategy developed in the subsequent subsections. 
\subsection{Background of DRL}
We consider a  reinforcement learning framework where an agent or controller dynamically interacts with an unknown environment, $\mathcal{E}$, by taking sequential decisions or actions in discrete time steps. At each time step, $t$, the agent interacting with the environment observes a state, $s_t \in \mathcal{S}$, selects an action, $a_t$, from a set of allowable actions, $\mathcal{A}$, and receives an immediate scalar reward, $r_t \in \mathcal{R}(s_t, a_t)$. Based on agent's current action, the agent enters into the new state, $s_{t+1}$. The cumulative discounted reward, $R_t$, at time step, $t$, is defined as 
\begin{align}
R_t = \sum_{k=0}^{\infty}\gamma^k r_{t+k},
\end{align}
where $\gamma \in (0,1]$ is the reward discount factor, which balances between the impact of recent rewards and earlier rewards. The learning objective is to maximize the expected cumulative reward at each state, $s_t$.
	
Q-learning is one of the most widely used training algorithms for reinforcement learning. It is a model-free off-policy technique where the agent, through interaction with the environment, estimates the expected return for taking an action while in a particular state. The Q-value, $Q^\pi (s,a)$, for state-action pair, $(s,a)$, is defined as the expected cumulative discounted reward for taking action, $a$,  in state, $s$, and following a policy, $\pi$, onward, i.e.,
\begin{align}\label{Q_pi}
Q^\pi (s,a)=\mathds{E}[R_t|s,a],
\end{align}
where $\mathds{E}[\cdot]$ denotes expectation. Q-learning adopts a value iteration approach to find the Q-values for each state-action pair, and optimal value function $Q^* (s,a)$ is the one which provides maximum action value for state, $s$, and action, $a$, achievable by following any policy:
\begin{align}
Q^* (s,a)=\max_{\substack{\pi}}Q^\pi (s,a).
\end{align}
Accordingly, optimal policy is given by
\begin{align}
\pi^* (s)=\text{argmax}_{\substack{a}}Q^* (s,a).
\end{align}
Using Bellman equation \cite{sutton2018reinforcement}, the value function in \eqref{Q_pi} can be unrolled recursively as
\begin{align}
Q^{\pi}(s,a)=\mathds{E}_{s'}\left[r_t+\gamma Q^{\pi}(s',a')|s,a \right].
\end{align}
Hence, the optimal value function can also be unrolled as 
	
\begin{align}
Q^*(s,a)=\mathds{E}_{s'}\left[r_t + \gamma \max_{\substack{a'}}Q^* (s',a')|s,a\right].
\end{align}
The value iteration algorithm can solve the Bellman equation, and the update rule is given by 
\begin{align}
Q_{i+1}(s,a)\leftarrow  \mathds{E}_{s'}\left[r_t + \gamma \max_{\substack{a'}}Q_i (s',a')|s,a.\right].
\end{align}
	
In deep Q-learning, the value functions are approximated by deep neural network parameterized by the weights, $\zeta$:
\begin{align}
Q(s,a,\zeta)\approx Q^\pi (s,a)
\end{align}
This helps to estimate the Q-values even for very large state-action space, and reduces the computational complexity.
%	\begin{align}
%	\mathcal{L}(w)=\mathds{E}\left[\left(r+\gamma \max_{\substack{a'}} Q (s',a',w)-Q(s,a,w)\right)^2\right]
%	\end{align}
%	
%	\begin{align}
%	\frac{\delta\mathcal{L}(w)}{\delta w}
%	\end{align}
\subsection{Beam Learning Framework}
Appropriate MIMO broadcast beam selection for cell-sectorization is critical for wireless network performance optimization. Our objective here is to build a mechanism that automatically facilitates the selection of best beams for all the sectors. Moreover, we would need the sectors autonomously update their beam parameters based on different scenarios or user distributions, and hence realize self-tuning sectorization. Towards this goal, our learning framework can be described as follows:

\textbf{Specification of design parameters:} First of all, network designer needs to decide on the objective function that needs to be optimized \cite{Self_Driving_Radios}. For broadcast beam optimization, an important objective function is the network coverage or total number of connected UEs in the network. The optimization parameters in this problem are the beam weights for each sector antenna element. It is necessary to select the optimal beam for each BS from a set of possible beams. Next, the system designer needs to decide on what input, such as RSRP or RSRQ, are required from the UEs in order to learn their mobility behavior and optimize the beams. Finally, in order to avoid random broadcast beams during the deployment stage, a simulation platform based on ray-tracing data is built to train the DRL agent offline.

\textbf{Learning Engine:}
An agent or learning engine has the task of learning the UE mobility pattern and selecting the best beam parameters for each scenario. It takes feedbacks from UEs as inputs, and suggests the optimal beam vectors for all sectors. Updating the beams based on user distribution by autonomously identifying underlying mobility pattern requires training. However, online training is often not desirable because of stringent network management requirements from the operators. Hence, the training needs to be done offline, and the training environment has to be close to the real cellular environment as much as possible so that the optimal beams in the training stage will be identical to the optimal beams in deployment stage-- the procedure is presented in details in the next subsection.

\textbf{Online Deployment and Occasional Re-training}
Once the learning engine is trained offline, the learned agent is deployed for real-time operation. It will enable the BSs to choose the optimal beams and update the selections based on users mobility pattern. Since users' mobility pattern in the network don't change too frequently, the beam parameters learned offline can remain unchanged for a long period of time-- on the order of weeks or months. Whenever, there is a need to support new scenarios or any change in mobility patterns is identified, the learning engine would need to be re-trained offline based on recent data. The newly learned beam parameters will then be pushed to the respective BSs for updated operation.

\subsection{Offline Training}
Dynamically updating the broadcast beam patterns according to the cellular environment and user distribution for all cells in real time is intrinsically a difficult problem. Directly deploying a DRL agent and training it online is not only slow but also costly. During the online training stage of the DRL, the agent may output some random beams according to the greedy exploration algorithm. Some of these random beams may not be acceptable to operators because of degraded network performance. In order to address this issue, we develop an offline training mechanism using ray-tracing data to train the DRL network before real deployment. By providing azimuth angle of arrival, elevation angle of arrival, azimuth angle of departure, elevation angle of departure, and path loss value of each path for each location in a cell, ray-tracing can well-capture the cellular environment so that the learned beam in the offline training platform could be the same as the online deployment case.
 The offline training is focused on learning the UE distribution pattern from users' location history data. The location data includes UEs' location and the corresponding time stamp. The location history data contains the UEs' mobility pattern information. Together with ray-tracing data, which contains the information about signal propagation environment , UEs' location history data are used to train the DRL network so that the DRL agent could learn the best broadcast beam according to both the cellular environment and UE distribution pattern. After offline training the DRL network, it will be deployed to provide real-time broadcast beam selection results for all the BSs in the cellular network. In the following, we describe the detailed steps of offline training. 
	
\begin{figure}[h!]
\centering
\includegraphics[width=1.0\linewidth]{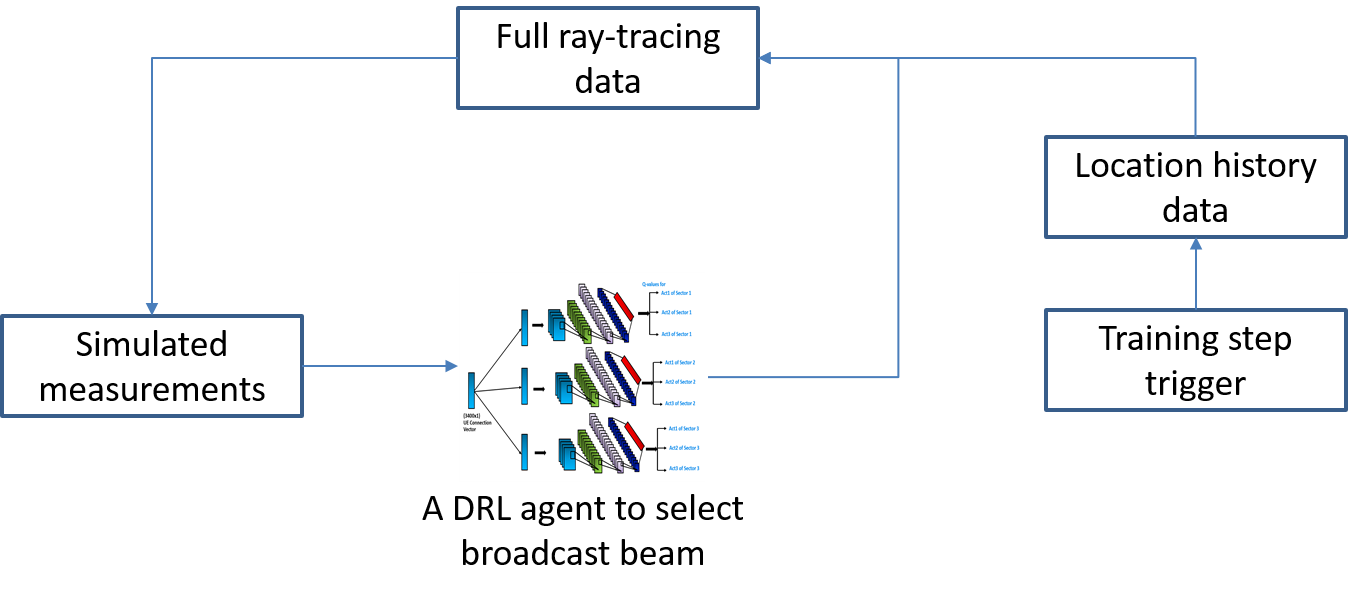}
\caption[width=0.5\linewidth]{{Offline training}}
\label{mobility}
\end{figure}

According to 3GPP standard on minimization of drive test(MDT), a BS could configure its UEs to report measurement results, time stamp, and location information~\cite{3GPP_TS_MDT}. Therefore, we assume that UE location history information is available for a cellular network. During one training step, a batch of time stamps are selected from the location history data, and the corresponding UEs' location information is incorporated to ray-tracing data for every time stamp. Therefore, the UE distribution at the selected timestamp is combined with ray-tracing data. We call the ray-tracing data with UE distribution information as scenario-specific ray-tracing data and the UEs who report their measurement information during the timestamp as selected UEs. Based on the current BSs' broadcast beam and scenario-specific ray-tracing data, the receive power for the selected UEs could be calculated and accordingly the network coverage. A reward could be provided to the DRL agent based on the coverage and the DRL agent could accordingly update its selection of broadcast beams based on selected optimizer. These offline training steps could be repeated many times until the DRL agent converges. After the DRL agent converges, it could be deployed in the cellular network for real-time broadcast beam selection. Details on the DRL agent design is discussed in next section. The entire offline training process is pictorially depicted in Fig.~\ref{mobility} and Algorithm~\ref{alg:offline_training}.

% \begin{figure}
%     \centering

% 	\begin{tikzpicture}[node distance=2cm]
	
% 	\node (start) [startstop] {Start};
% 	\node (dec1) [decision, below of=start, yshift=-1.3cm] {Algorithm Converge?};
% 	\node (stop) [startstop, right of=dec1, xshift=3.0cm] {Stop};
% 	\node (pro1) [process, below of=dec1, yshift=-1.5cm] {Select the UE distribution according to history data};
% 	\node (pro2) [process, below of=pro1, yshift=-0.0cm] {Select the best beam for sectors according to Q-values};
% 	\node (pro3) [process, below of=pro2, yshift=-0.2cm] {UE conducts measurement according to scenario-specific ray-tracing data; a reward is obtained as a function of all the measurement data};
% 	\node (pro4) [process, below of=pro3, yshift=-0.2cm] {DRL updates its weights based on the learning algorithm and reward};
	
% 	\draw [arrow] (start) -- (dec1);
% 	\draw [arrow] (dec1)-- node[anchor=east] {No}(pro1);
% 	\draw [arrow] (pro1) -- (pro2);
% 	\draw [arrow] (pro2) -- (pro3);
% 	\draw [arrow] (pro3) -- (pro4);
% 	\draw [arrow] (dec1) -- node[anchor=south] {Yes} (stop);
% 	\draw [arrow] (pro4.west) -- ++(-2.6,0) |-  (dec1.west);
% 	\end{tikzpicture}
% 	 \caption{Algorithm Flow for Offline Training.}
% 	 \label{Learning_Flow}
% 	\end{figure}
\begin{algorithm}[!htbp] 
    \small
	\caption{ Offline Training} 
	\label{alg:offline_training} 
	\begin{algorithmic}[1] 
		\Require
		\State UE location history data, ray-tracing data of a cellular network
		\Ensure
		\State trained DRL agent for broadcast beam selection
		\INITIALIZATION 
		\State Define a pool of candidate antenna patterns; 
		 \LEARN
		 \While{algorithm doesn't achieve convergence}
		 \State Select a batch of UE location at different timestamps;
		 \State incorporate UE location distribution to ray-tracing data to create scenario-specific ray-tracing data;
		 \State calculate the received power for each UE in the scenario-specific ray-tracing data based on the current BSs' broadcast beam;
		 \State calculate the network coverage, and calculate a total reward as a function of network coverage;
		 \State DRL updates its neural weights based on the learning algorithm and reward
		 \EndWhile
		 
% 		\algstore{myalg}
    \end{algorithmic}
\end{algorithm}

% 	\begin{figure}[h!]
% 		\centering
% 		\includegraphics[width=0.8\linewidth]{Figures/Learning_Flow}
% 		%\captionsetup{margin= {20pt},justification=centerlast,skip=-5pt,font=normalsize}
% 		%\vspace{0.1cm}
% 		\caption[width=0.5\linewidth]{{Learning Flow for Offline Training.}}
% 		%\captionsetup{justification=centering}
% 		\label{Learning_Flow}
% 	\end{figure}

	\section{DRL for Broadcast Beam Optimization}\label{DRL_optimization}
	
\begin{figure}[h!]
		\centering
		\includegraphics[width=0.8\linewidth]{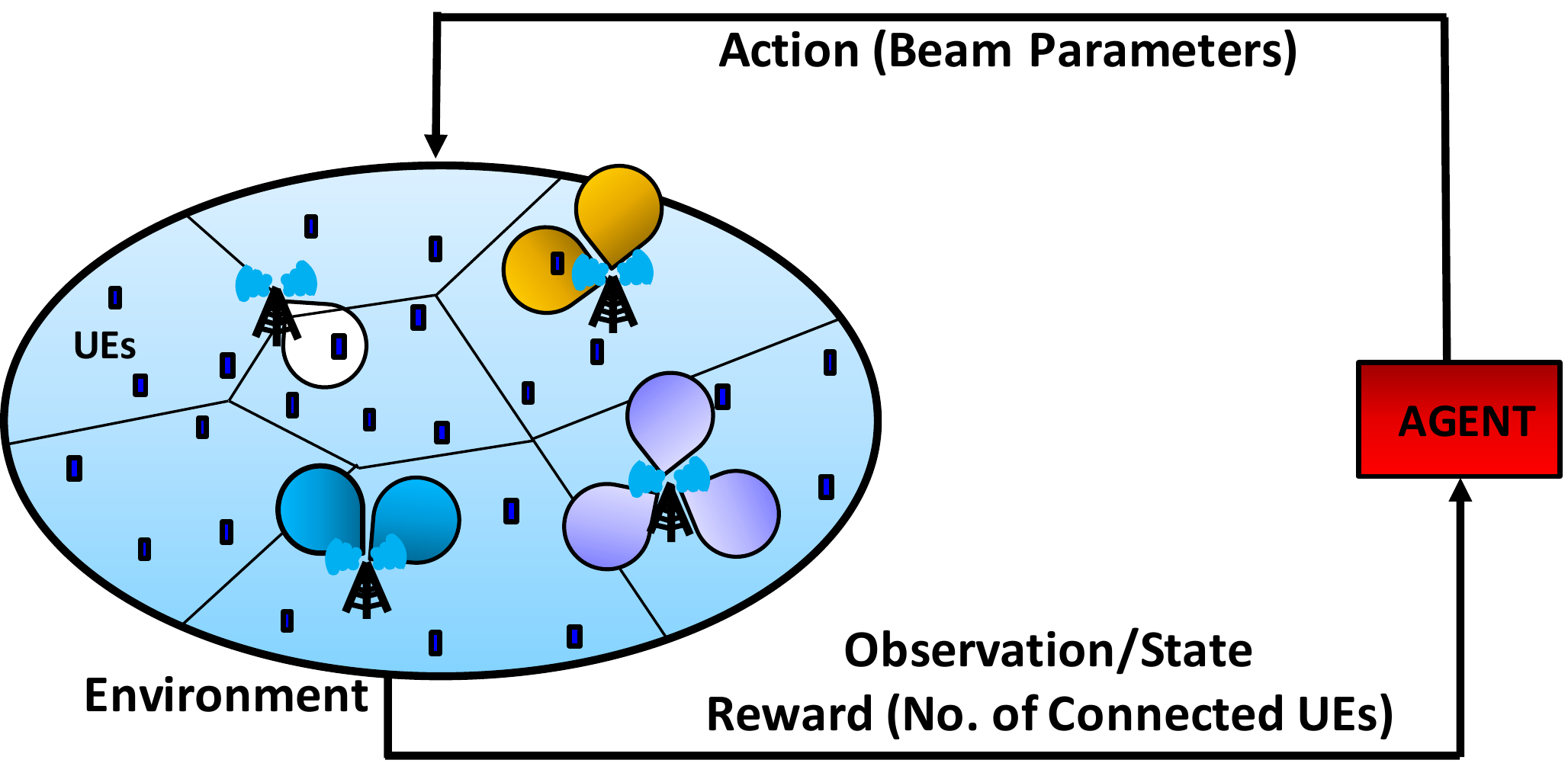}
		%\captionsetup{margin= {20pt},justification=centerlast,skip=-5pt,font=normalsize}
		%\vspace{0.1cm}
		\caption[width=0.5\linewidth]{{Reinforcement Learning Framework for Beam Optimization}}
		%\captionsetup{justification=centering}
		\label{framework}
	\end{figure}

	In this section, details on the design of DRL framework for self-tuning sectorization are presented. The DRL network is utilized in order to track  optimal beams during both the offline training and online deployment.  To be specific, a deep Q-network (DQN)-based architecture has been proposed to select MIMO broadcast beams for all sectors in a dynamic environment. For better stability of the results, we use DQN with \textit{experience replay} \cite{lin1993reinforcement, mnih2013playing}.
	The agent (decision maker) interacts with the environment by selecting the best broadcast beam parameters. The DRL has three main components: state, action and reward. The dynamics between state, action and reward are shown in Fig.~\ref{framework}. The agent interacts with the environment by observing the state of the network, and taking action that maximizes the reward or network performance metric. Next we describe each of these components in details, and explain how we model the state, action, and reward in DRL-based MIMO broadcast beam optimization problem.
	
\begin{figure}[h!]
		\centering
		\includegraphics[width=1.0\linewidth]{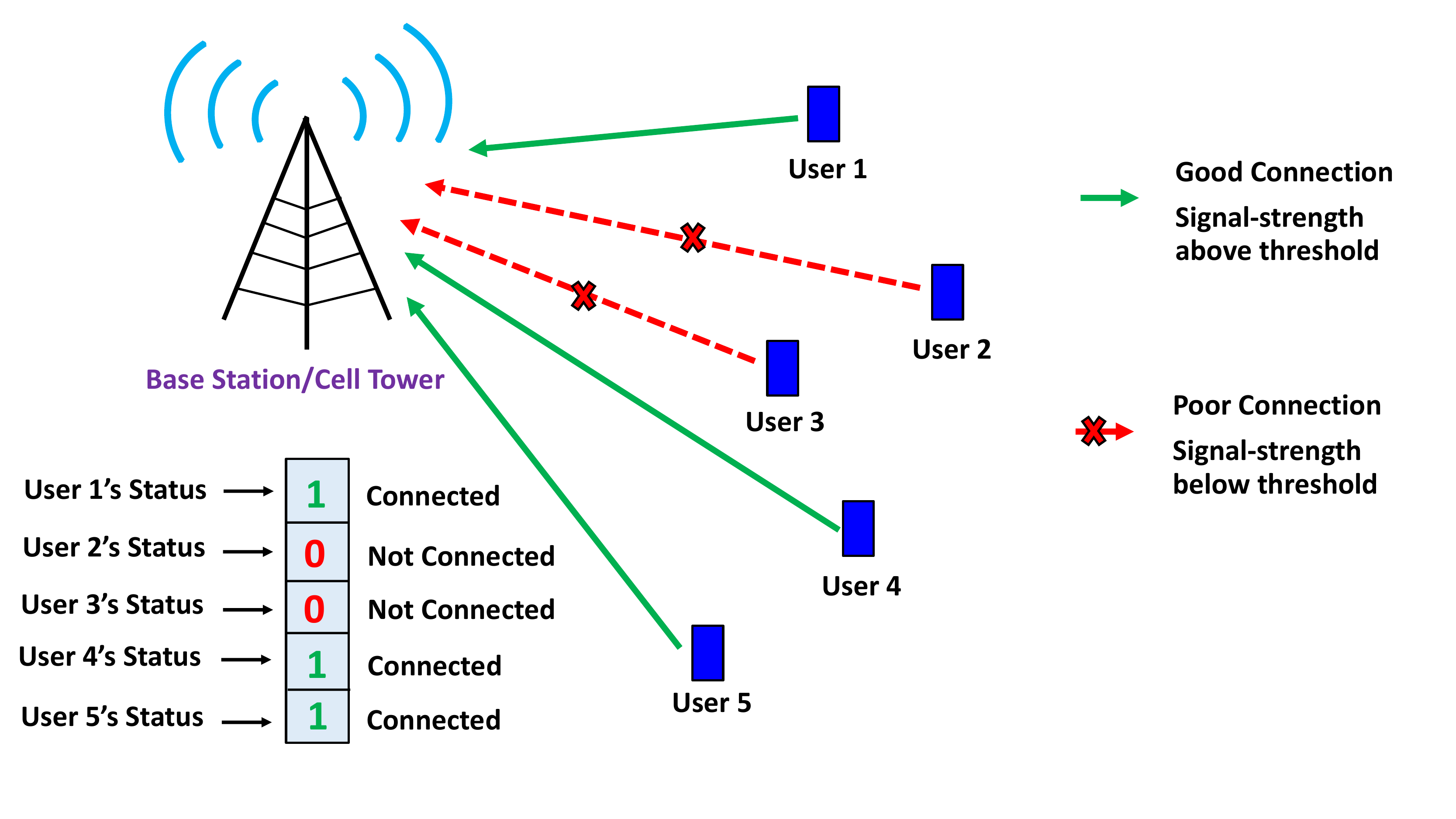}
		%\captionsetup{margin= {20pt},justification=centerlast,skip=-5pt,font=normalsize}
		%\vspace{0.1cm}
		\caption[width=0.5\linewidth]{{DRL \text{State} Representation for Beam Optimization Problem}}
		%\captionsetup{justification=centering}
		\label{state}
	\end{figure}

	\textbf{State}: \textit{State} in the proposed RL framework is designed as to reflect the network coverage situation which can be obtained from UE measurements. To be specific, we can design the state as the connection indicators of UEs in the network (a vector of 1/0s). Each UE reports its status to its attached BS. If a UE's SINR falls below a predefined threshold, $T$, a zero is placed at the element of the vector corresponding to that UE. Otherwise, a one is placed. Accordingly, a ‘0’ in the state vector will represent that the corresponding UE has poor connection, and a ‘1’ will indicate that the UE has good connection. 
%	Another method to design the state could be a vector of the RSSI, RSRQ, RSRP or SINR. 
	The DRL state representation adopted in this work is pictorially depicted in Fig.~\ref{state}.
	
	\textbf{Action}: An \textit{action} of the agent is defined as the selection of beam index from a pool of candidate beam patterns. The agent observes the states and the corresponding reward, and takes the best possible action that maximizes the cumulative discounted future reward for the next time step. At the beginning of the training, the agent explores different actions in an attempt to learn the best beams for different user distribution. However, once the training phase is complete, the agent exploits the learned information and only selects the best known actions that maximize the cumulative reward for each user distribution.
	
	\textbf{Reward}: A \textit{reward} in this work refers to any network performance metric. One way to design the reward can be the total number of connected UEs in the network based on the state and action taken in the previous state. Another approach to design the reward can be the function of the measurement results, for example, a function of the SINR or RSRP vector. In this work, we adopt the first approach for designing the reward. It is to be noted here that maximizing  total number of connected UEs in the network is equivalent to maximizing the coverage of the cellular network.

	The agent's goal is to maximize the cumulative discounted future reward. The agent gathers its experiences as tuples, $(s_t, a_t, r_t, s_{t+1})$, where $s_t$ denotes  current UE connection state, $a_t$  denotes the action taken at state, $s_t$; $r_t$  is the instantaneous reward obtained from state, $s_t$ and by taking action, $a_t$; and $s_{t+1}$  is the next state. The agent stores history of its experiences in a memory called  experience replay memory\cite{lin1993reinforcement}, and replay memory stores the tuples, $(s_t, a_t, r_t, s_{t+1})$, for all time steps. The DRL agent randomly samples mini-batches of experience from the replay memory for training, and selects an action based on $\epsilon$-greedy policy, i.e., with probability $\epsilon$, it tries a random action, and with probability $(1-\epsilon)$, the agent selects the best known action so far. The optimum action in a particular state is selected based on maximum Q-values \cite{sutton2018reinforcement} corresponding to that state. In DQN-based reinforcement learning, the Q-values are predicted using deep neural network. Input to the neural network is the UEs' connection vector representing the state of the RL environment, and output is the Q-values corresponding to all the possible actions, i.e., beam indices from the beam-pool. In the following subsections, we detail the broadcast beam optimization strategies for both single cell and multiple cell scenarios. 
	
%	\subsection{State-vector Transformation}

	\subsection{Broadcast beam optimization for single cell environment}
	
\begin{figure}[h!]
		\centering
		\includegraphics[width=0.6\linewidth]{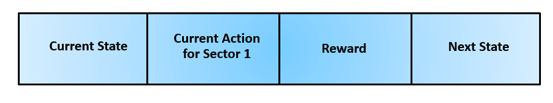}
		%\captionsetup{margin= {20pt},justification=centerlast,skip=-5pt,font=normalsize}
		%\vspace{0.1cm}
		\caption[width=0.5\linewidth]{{Replay Memory for Single Cell Beam Optimization}}
		%\captionsetup{justification=centering}
		\label{replay_single}
	\end{figure}

	In the single cell case,  beam parameters corresponding to only one sector need to be optimized. This could serve as an example where a legacy LTE sector is replaced with one massive MIMO unit. 
%	\begin{figure}[h!]
%		\centering
%		\includegraphics[width=0.8\linewidth]{Figures/single_sector}
%		%\captionsetup{margin= {20pt},justification=centerlast,skip=-5pt,font=normalsize}
%		%\vspace{0.1cm}
%		\caption[width=0.5\linewidth]{{Single Sector Beam-synthesis}}
%		%\captionsetup{justification=centering}
%		\label{single_sector}
%	\end{figure}
	The goal is to maximize the number of connected UEs for different dynamic user distributions. The agent keeps a single replay memory containing the agent's experience tuples, and randomly samples from it--this random sampling from experience replay memory helps to decorrelate the data\cite{mnih2015human}. The replay memory architecture for single sector case is shown in Fig.~\ref{replay_single}. Next, we describe the the neural network architecture for Q-value prediction for single cell scenario.
	
\begin{figure}[h!]
		\centering
		\includegraphics[width=1.0\linewidth]{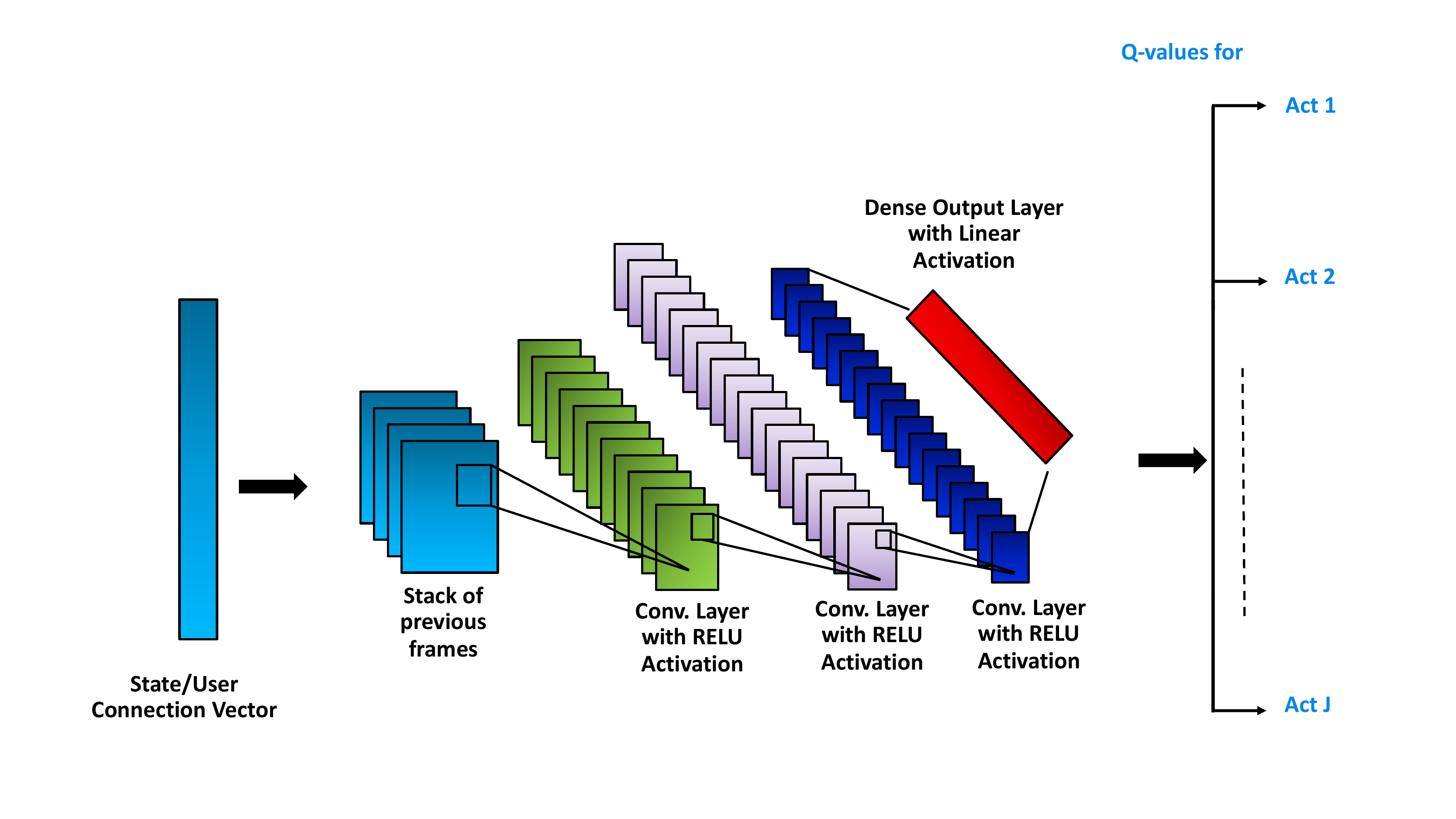}
		%\captionsetup{margin= {20pt},justification=centerlast,skip=-5pt,font=normalsize}
		%\vspace{0.1cm}
		\caption[width=0.5\linewidth]{{Neural Network Architecture for Single Sector Beam Optimization}}
		%\captionsetup{justification=centering}
		\label{architecture_single}
	\end{figure}

	\textbf{Neural Network architecture for single sector case:} For Q-value prediction, a deep convolutional neural network is used in this work. 
	For the suitability of computing the Q-values using convolutional neural network, we transform the $(K \times 1)$ UE connection vector into an $(\frac{K}{100} \times 100)$ frame. Four such frames are stacked together, and fed as the input to the neural network for computing the Q-values. 
%	For clarity, the state-vector transformation process is shown in Fig.~\ref{state_transformation}.
%		\begin{figure}[h!]
%		\centering
%		\includegraphics[width=0.8\linewidth]{Figures/state_transformation}
%		%\captionsetup{margin= {20pt},justification=centerlast,skip=-5pt,font=normalsize}
%		%\vspace{0.1cm}
%		\caption[width=0.5\linewidth]{{State-vector Transformation}}
%		%\captionsetup{justification=centering}
%		\label{state_transformation}
%	\end{figure}
	We used three convolutional layers--all with rectified linear unit (ReLU) activation function. First convolution layer has 32 (8x8) filters. Second and third convolution layers have 64 (4x4) filters and 64 (3x3) filters, respectively. Finally, a dense layer with linear activation function is used as the output layer. The neural network architecture for predicting the Q-values for single sector case is shown in Fig.~\ref{architecture_single}.
	Two such identical neural networks are used in predicting the Q-values. One is used for computing the running Q-values--this neural network is called the \textit{evaluation network}. The other neural network, called the \textit{target network}, is held fixed for some training duration, say for $P$ episodes, and every $P$  episodes, the weights of the evaluation neural network is transferred to the target neural network. It has been shown that this two neural network-based approach for Q-value prediction provides better stability of results at convergence \cite{mnih2015human}.
%	The weight transfer of the neural network is depicted in Fig.~\ref{DQN_single}:
%	\begin{figure}[h!]
%		\centering
%		\includegraphics[width=1.0\linewidth]{Figures/DQN_single}
%		%\captionsetup{margin= {20pt},justification=centerlast,skip=-5pt,font=normalsize}
%		%\vspace{0.1cm}
%		\caption[width=0.5\linewidth]{{DQN Learning Structure}}
%		%\captionsetup{justification=centering}
%		\label{DQN_single}
%	\end{figure}
%	If the instantaneous reward, i.e., the number of connected UEs at time step, $t$,  is denoted as $r_t$, then the total discounted future reward at time step, $t$, is defined as 
%	\begin{align}
%	R_t=\sum_{t'=t}^{T}\gamma^{t'-t}r_t
%	\end{align}
%	where $\gamma$ is the reward discount factor, and  denotes the episode duration. 
%	\textbf{State-vector Transformation example}: For the suitability of computing the Q-values using convolutional neural network, which is used for predicting the Q-values in this work, we transform the (3400 x 1) user connection vector into a (34 x 100) frame. Four such frames are stacked together, and fed as the input to the neural network. For clarity, the transformation is shown in Fig.~\ref{state_transformation_eg}.
%	\begin{figure}[h!]
%		\centering
%		\includegraphics[width=0.8\linewidth]{Figures/state_transformation_eg}
%		%\captionsetup{margin= {20pt},justification=centerlast,skip=-5pt,font=normalsize}
%		%\vspace{0.1cm}
%		\caption[width=0.5\linewidth]{{Example of State-vector transformation}}
%		%\captionsetup{justification=centering}
%		\label{state_transformation_eg}
%	\end{figure}
	The detailed procedure for broadcast beam optimization for single sector case is summarized  in Algorithm~\ref{alg:single_sector}. 
	
%	\begin{algorithm}
%		\caption{Boradcast Beam Optimization for Single Sector}\label{alg:single}
%		\begin{algorithmic}[1]
%			INPUT
%			\Procedure{Euclid}{$a,b$}\Comment{The g.c.d. of a and b}
%			\State $r\gets a\bmod b$
%			\While{$r\not=0$}\Comment{We have the answer if r is 0}
%			\State $a\gets b$
%			\State $b\gets r$
%			\State $r\gets a\bmod b$
%			\EndWhile\label{euclidendwhile}
%			\State \textbf{return} $b$\Comment{The gcd is b}
%			\EndProcedure
%		\end{algorithmic}
%	\end{algorithm}

\begin{algorithm}[!htbp] 
    \small
	\caption{ Broadcast Beam Optimization for Single Sector} 
	\label{alg:single_sector} 
	\begin{algorithmic}[1] 
		\Require
		\State RSRP measurements from the UEs in the network
		\Ensure
		\State Optimum broadcast beam pattern that maximizes the number of connected UEs
		\INITIALIZATION 
		\State Define the pool of candidate antenna pattern; 
		\State Define the maximum exploration rate, $\epsilon_{max}$, minimum exploration rate, $\epsilon_{min}$, exploration decay rate, optimizer's learning rate, $\alpha$, and reward discount factor, $\gamma$;
		\State Initialize the replay memory, $\mathcal{D}$.
		 \OPTIMIZATION
		 \For{$ \text{episode} = 1, 2, \ldots,  Z, $}
		\State Initialize the state vector at time step 1 as $\mathbf{s}_1$;
		\For{$ t = 1,2, \ldots, T', $}
		\State Sample $c$ from Uniform $(0,1)$
		\If{$c \leq \epsilon$}
		\State Select an action (choose a beam index) randomly from the pool of action set (candidate beam indices)
		\Else
		\State Select an action, $a_t=\text{argmax}_{\substack{a}}$ $Q^*(\mathbf{s}_t,a,\theta)$
		\EndIf
		\State Apply the selected beam weights on the antenna array. BS transmits sector-specific signals using the newly selected antenna weights
		\State BS receives UE measurements and report the measurements to the agent. Observe the resulting RL state, $\mathbf{s}_{t+1}$, which is the UE connection vector.
		\State Pre-process the state vector into a frame before feeding as  input to Neural Network
		\State Compute the reward, $r_t$, which is the number of connected UEs.
		\State Store the experience tuple, $e_t=(\mathbf{s}_t, a_t, r_t, \mathbf{s}_{t+1}),$ in replay memory, $\mathcal{D}$
		\State Sample random mini-batches of experience $(\mathbf{s}_j, a_j, r_j, \mathbf{s}_{j+1})$, from $\mathcal{D}$
		
		\If{$s_{j+1}$ is a terminal state}
		\State Set $y_j=r_j$
		\Else
		\State Set $y_j=r_j+\gamma\max_{\substack{a'}} Q'(\mathbf{s}_j, a'; \theta)$
		
		\EndIf
		\State Perform a gradient descend on $\left(y_j-Q(\mathbf{s}_j, a_j; \theta)\right)^2$
		\EndFor
	
		\EndFor
% 		\algstore{myalg}
		\end{algorithmic}
		\end{algorithm}
		
% 		\begin{algorithm}                     
% 		\begin{algorithmic} [1]                   % enter the algorithmic environment
% 		\algrestore{myalg}

% 		\If{$s_{j+1}$ is a terminal state}
% 		\State Set $y_j=r_j$
% 		\Else
% 		\State Set $y_j=r_j+\gamma\max_{\substack{a'}} Q'(\mathbf{s}_t, a'; \theta)$
		
% 		\EndIf
% 		\State Perform a gradient descend on $\left(y_j-Q(s_j, a_j; \theta)\right)^2$
% 		\EndFor
	
% 		\EndFor

% 	\end{algorithmic} 
% \end{algorithm}

%	\begin{algorithm}[H]
%		\SetAlgoLined
%		\KwResult{Write here the result }
%		initialization\;
%		\While{While condition}{
%			instructions\;
%			\eIf{condition}{
%				instructions1\;
%				instructions2\;
%			}{
%				instructions3\;
%			}
%		}
%		\caption{How to write algorithms}
%	\end{algorithm}
%%
%%	
%	\begin{algorithm}[H]
%		\KwData{this text}
%		\KwResult{how to write algorithm with \LaTeX2e }
%		initialization\;
%		\While{not at end of this document}{
%			read current\;
%			\eIf{understand}{
%				go to next section\;
%				current section becomes this one\;
%			}{
%				go back to the beginning of current section\;
%			}
%		}
%		\caption{How to write algorithms}
%	\end{algorithm}

%	\begin{figure}[h!]
%		\centering
%		\includegraphics[width=0.8\linewidth]{Figures/Algorithm_1}
%		%\captionsetup{margin= {20pt},justification=centerlast,skip=-5pt,font=normalsize}
%		%\vspace{0.1cm}
%		\caption[width=0.5\linewidth]{{Algorithm-1}}
%		%\captionsetup{justification=centering}
%		\label{Algorithm_1}
%	\end{figure}
	\subsection{Broadcast Beam Optimization for multiple cell environment:}
	
	In this subsection, we present the framework for dynamically optimizing MIMO broadcast beams for multiple sector environment, where the RL agent needs to simultaneously control the beam parameters for all the sectors based on different user distributions. For this case, there needs to be some  significant changes on the RL framework compared to that for single sector beam optimization. In the multiple  sector environment, each sector has its own pool of beams or action sets. Each sector can hence independently select its own beam parameters. 	The setup is similar to that of multi agent system \cite{sycara1998multiagent, wooldridge2009introduction}. The goal remains  the same--to maximize  the overall network coverage. This is a challenging problem in terms of computational tractability. For an illustration, let us consider that there are $m$ sectors in the network, and each sector has $j$ possible beam patterns (actions) to select from. Hence, total number of actions, i.e., all possible combinations of sectors' beam patterns, becomes $j^m$, which increases exponentially with  total number of sectors. If there are $40$ base stations, and each has $5$ possible actions to choose from, total possible combination of beam patterns becomes $5^{40}$, which is an  extraordinarily large number, making it difficult to achieve optimal solution within reasonable time.
	
	One way to find the appropriate broadcast beams for multiple sectors simultaneously can be simply to extend the single sector framework developed in the previous subsection. In other word, a single large neural network with large number of output nodes can be used to predict the Q-values for all possible $j^m$ actions. 
However, total number of training samples needed to train such neural network would be extremely large, which may not be feasible at all for any practical purposes. In other words, the learning algorithm can almost never achieve convergence with this architecture for even moderate size cellular network.

To address this issue,  we propose a novel low-complexity algorithm for optimizing the broadcast beams for multiple sectors where the action space grows only linearly, instead of exponentially, with total number of sectors in the network. 
	Let us again assume that there are $m$ sectors, and each sector has $j$ possible actions (beam-weight set) to choose from. Unlike the single cell case, for multiple cell environment, we assume  the agent  preserves  different replay memories for different sectors. Moreover we use $m$ different neural networks for independently computing the Q-values for $j$ sectors. Each neural network is responsible to predict the optimum action for the corresponding sector only. With this architecture, number of actions increases only linearly, but we can still achieve perfect convergence with reasonably short computation time, which demonstrated through extensive simulation in Section \ref{Simulation_section}. The details of the architectures for replay memory and neural networks for multiple sector broadcast beam optimization are briefly described next.
	
		\begin{figure}[h!]
		\centering
		\includegraphics[width=0.8\linewidth]{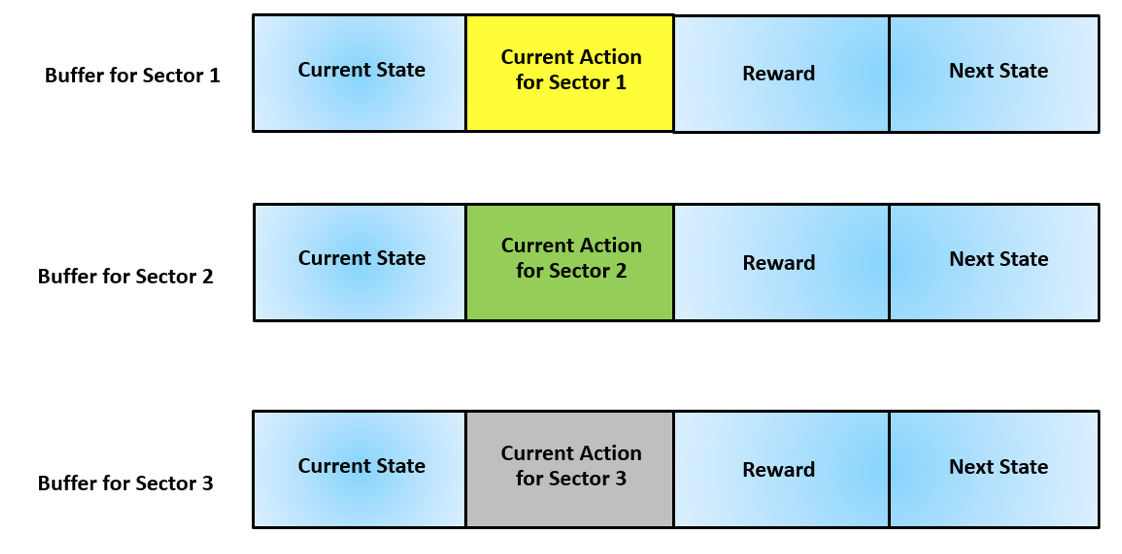}
		%\captionsetup{margin= {20pt},justification=centerlast,skip=-5pt,font=normalsize}
		%\vspace{0.1cm}
		\caption[width=0.5\linewidth]{{Replay Buffer architecture for multiple sector case}}
		%\captionsetup{justification=centering}
		\label{memory_multiple}
	\end{figure}

	\textbf{Replay memory architecture:} The replay memory architecture for multiple sectors broadcast beam optimization is shown in figure \ref{memory_multiple}. There are separate buffers for each sector. The same current state, reward, and the next state are stored in all the replay memories/buffers for the sectors. However, the replay memories differ in the actions taken (beam indices chosen) by the each sector. While all the sectors observe the same current state, $s_t$ , reward, $r_t$ , and next state, $s_{t+1}$ , the action stored are different--BS 1's action is stored in buffer 1, BS 2's action is stored in buffer 2, and so on. The rationale behind this buffer architecture is that states and rewards are network specific, and same states and rewards are observed by all sectors. On the other hand, each sector takes its own action, and their joint actions regulate the overall network state and the corresponding reward.
	
\begin{figure}[h!]
		\centering
		\includegraphics[width=1.0\linewidth]{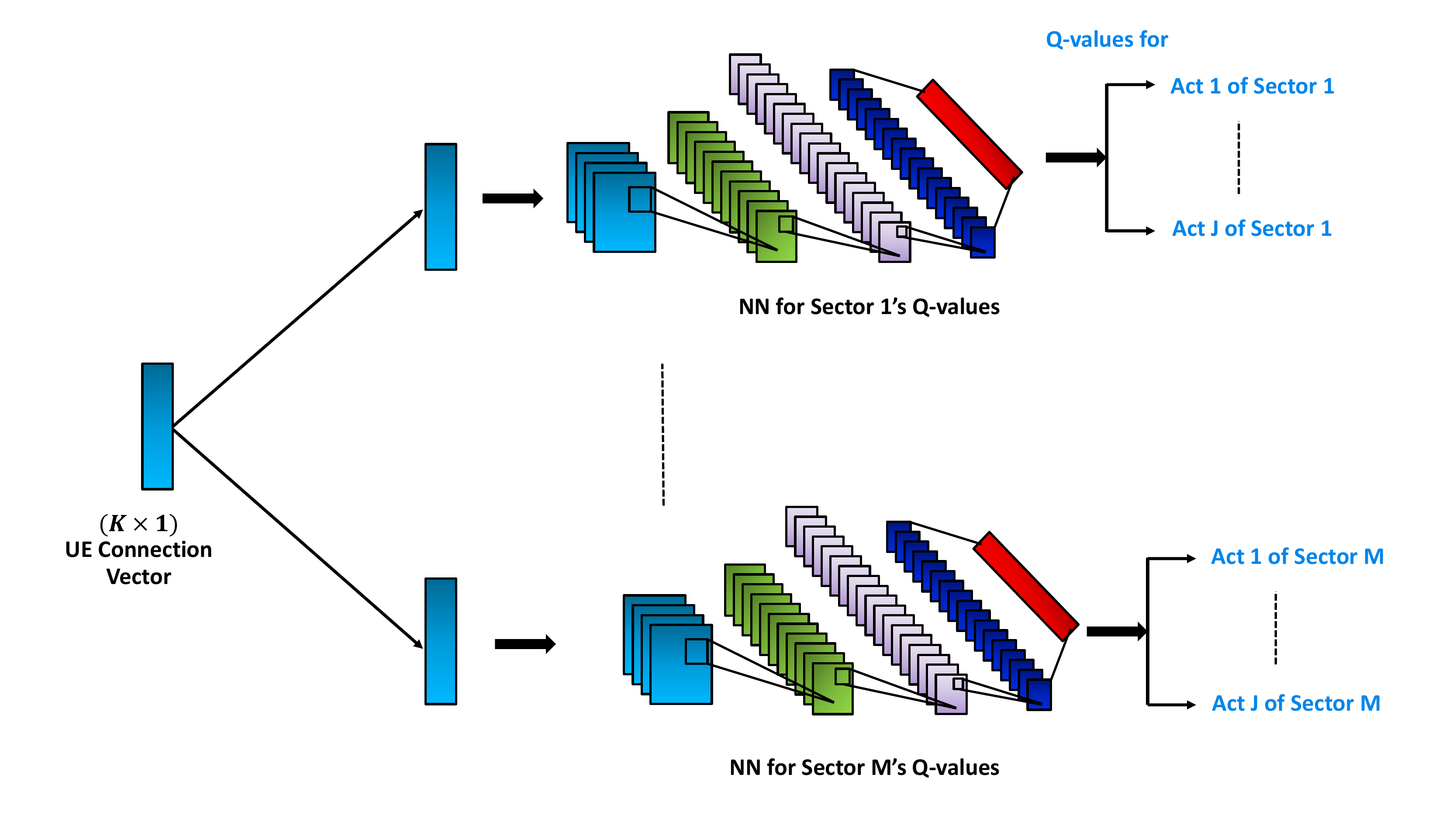}
		%\captionsetup{margin= {20pt},justification=centerlast,skip=-5pt,font=normalsize}
		%\vspace{0.1cm}
		\caption[width=0.5\linewidth]{{Neural Network architecture for multiple sector case}}
		%\captionsetup{justification=centering}
		\label{NN_multiple}
	\end{figure}

	\textbf{Neural Network architecture:} The neural network architecture for predicting the Q-values for multiple sectors are shown in Fig.~ \ref{NN_multiple}. 
	The depiction is presented for $M$ sectors case, where $M$ separate neural networks are used for predicting the Q-values for $M$ sectors. Input to all neural networks are the same state vectors. Neural networks are identical, and the number of output for each neural networks is $J$. Hence, size of action space is $JM$, instead of $J^M$, i.e., total number of actions grows only linearly with number of sectors. The optimal action predicted by the Q-values of neural network 1 is stored in Buffer 1, which corresponds to sector 1. Similarly, the action predicted by the Q-values of neural network 2 is stored in Buffer 2, which corresponds to sector 2, and so on. The beam learning procedure for multiple BS environment is presented in Algorithm~\ref{alg:multiple_sector}.
	
	\begin{algorithm}[!htbp] 
	    \small
		\caption{ Broadcast Beam Optimization for Multiple Sectors} 
		\label{alg:multiple_sector} 
		\begin{algorithmic}[1] 
			\Require
			\State RSRP measurements from the UEs in the network
			\Ensure
			\State Optimum broadcast beam patterns for all sectors that maximizes the number of connected UEs
			\INITIALIZATION 
			\State Define a pool of candidate antenna pattern; 
			\State Define the maximum exploration rate, $\epsilon_{max}$, minimum exploration rate, $\epsilon_{min}$, exploration decay rate, optimizer's learning rate, $\alpha$, and reward discount factor, $\gamma$;
			\State Initialize the replay memory, $\mathcal{D}$.
			\OPTIMIZATION
			\For{$ \text{episode} = 1, 2, \ldots,  Z, $}
			\State Initialize the state vector at time step 1 as $\mathbf{s}_1$;
			\For{$ t = 1,2, \ldots, T', $}
			\State Sample $c$ from Uniform $(0,1)$
			\If{$c \leq \epsilon$}
			\State Select an action (choose a beam index) for each sector randomly from the beam pool
			\Else
			\For{$m=1,2, \ldots, M$}
			
			\State Select the action for $m$-th BS, $a_t^m=\text{argmax}_{\substack{a^m}} Q_m^*(\mathbf{s}_t, a^m; \theta_m)$

			\EndFor
			\EndIf

			\State Apply the selected beam patterns on the antenna arrays of the corresponding BSs
			
			\State  Observe the resulting RL state, $\mathbf{s}_{t+1}$, the UE connection vector.
			\State Pre-process the state vector into a frame before feeding to Neural Network
			\State Compute the reward, $r_t$, which is the number of connected UEs.
			
			\For{$m=1,2, \ldots, M$}
			\State Store the experience tuple for $m$-th sector, $e_t^m=(\mathbf{s}_t, a_t^m, r_t, \mathbf{s}_{t+1})$, in $m$-th replay memory, $\mathcal{D}^m$. 
			\State Sample random mini-batches of experience $(\mathbf{s}_j, a_j^m, r_j, \mathbf{s}_{j+1})$, from $\mathcal{D}^m$
			
			\If{$s_{j+1}$ is a terminal state}
			\State Set $y_j^m=r_j$
			\Else
			\State Set $y_j^m=r_j+\gamma\max_{\substack{a_m}} Q_m(\mathbf{s}_t, a_m; \theta)$
			
			\EndIf
			\State Perform a gradient descend on $\left(y_j^m-Q(s_j, a_j^m; \theta)\right)^2$
			\EndFor
			\EndFor
			
			\EndFor
			
% 			\algstore{myalg}
		\end{algorithmic}
	\end{algorithm}
	
% 	\begin{algorithm}                     
% 		\begin{algorithmic} [1]                   % enter the algorithmic environment
% 			\algrestore{myalg}

% 			\If{$s_{j+1}$ is a terminal state}
% 			\State Set $y_j^m=r_j$
% 			\Else
% 			\State Set $y_j^m=r_j+\gamma\max_{\substack{a_m}} Q_m(\mathbf{s}_t, a_m; \theta)$
			
% 			\EndIf
% 			\State Perform a gradient descend on $\left(y_j^m-Q(s_j, a_j^m; \theta)\right)^2$
% 			\EndFor
% 			\EndFor
			
% 			\EndFor
			
% 		\end{algorithmic} 
% 	\end{algorithm}

%	\begin{figure}[h!]
%		\centering
%		\includegraphics[width=1.0\linewidth]{Figures/Algorithm2}
%		%\captionsetup{margin= {20pt},justification=centerlast,skip=-5pt,font=normalsize}
%		%\vspace{0.1cm}
%		\caption[width=0.5\linewidth]{{Algorithm for Wide Beam Synthesis for Multiple Sector Case}}
%		%\captionsetup{justification=centering}
%		\label{Algorithm2}
%	\end{figure}

	\section{Simulation Results and Performance Analysis}\label{Simulation_section}
	In this section, we present the simulation results and performance evaluation for self-tuning sectorization mechanism through DRL-based MIMO broadcast beam optimization. We first present the results for single sector environment followed by multiple sectors case. Both periodic and Markov mobility patterns have been considered for the evaluation.

\begin{table}[ht]
    \caption{Simulation Parameters}
    \centering
    \begin{tabular}{c c}
    \hline\hline
    RL Parameter & Specification \\ [0.5ex] % inserts table %heading
    \hline
    Reward Discount Factor, $\gamma$ & 0.0001 \\
    Learning Rate, $\alpha$ & 0.001 \\
    Initial exploration probability, $\epsilon_{\text{max}}$& 1.0\\
    Final exploration probability, $\epsilon_{\text{min}}$& 0.000001\\
    Training batch size & 32\\
    Optimizer & Adam\\ 
    \hline\hline
    Network Parameter & Specification \\ [0.5ex] % inserts table %heading
    \hline
    Antenna array at BSs & $4 \times 4$ \\
    Antenna separation in azimuth domain & 1.48$\lambda$ \\
    Antenna separation in elevation domain & 0.5$\lambda$ \\
    UEs' SINR threshold for connectivit, $T$ & -6 dB\\
    BSs height from the ground & 35m\\
    \hline
    \end{tabular}
    \label{table:SimulationParameters}
\end{table}

	\subsection{ Results for single sector dynamic environment: }
	\label{simulation_singlecell}
\begin{figure}[h!]
		\centering
		\includegraphics[width=0.8\linewidth]{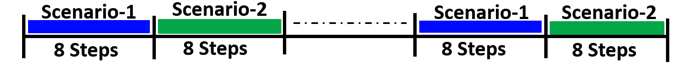}
		%\captionsetup{margin= {20pt},justification=centerlast,skip=-5pt,font=normalsize}
		%\vspace{0.1cm}
		\caption[width=0.5\linewidth]{{Periodic Change in Scenarios}}
		%\captionsetup{justification=centering}
		\label{scenario_pattern}
	\end{figure}

	In this sub-section, we present the performance evaluation for our proposed algorithm for single sector dynamic environment. The sector is equipped with a two dimensional (2D) antenna array with 4 antenna elements in both elevation and azimuth directions. The horizontal distance between BS antenna elements is 0.5 wave-length and the vertical distance between antenna elements is 1.48 wave-length. We first consider two scenarios or user distributions, and assume that users switch between Scenario-1 and Scenario-2 periodically every 8 time steps (see Fig.~\ref{scenario_pattern}). The BS is located at a height of 35 m from ground, and users are distributed randomly in the cell. Based on users' X-, Y-, and Z-coordinates, two scenarios are defined as follows: \textbf{Scenario-1}: $X \geq 2600 \text{ m}$, $Z\ge10 \text{ m}$; \textbf{Scenario-2}: $X \leq 2700 \text{ m}$, $Z\le 12 \text{ m}$. For simulation, this partition is used as users' mobility pattern. 
	The received power of each UE is calculated based on ray-tracing data. Noise level is set as $-95$ dBm, and SINR threshold level is kept at $-6$ dB. For a particular user, if the received SINR  is above this threshold, we consider the user to be connected;  otherwise, we consider it to be not-connected. A set of simulation parameters used in this work summarized in Table~\ref{table:SimulationParameters}.

\begin{figure}
		\centering
		\begin{subfigure}{.5\textwidth}
			\centering
			\includegraphics[width=1.0\linewidth]{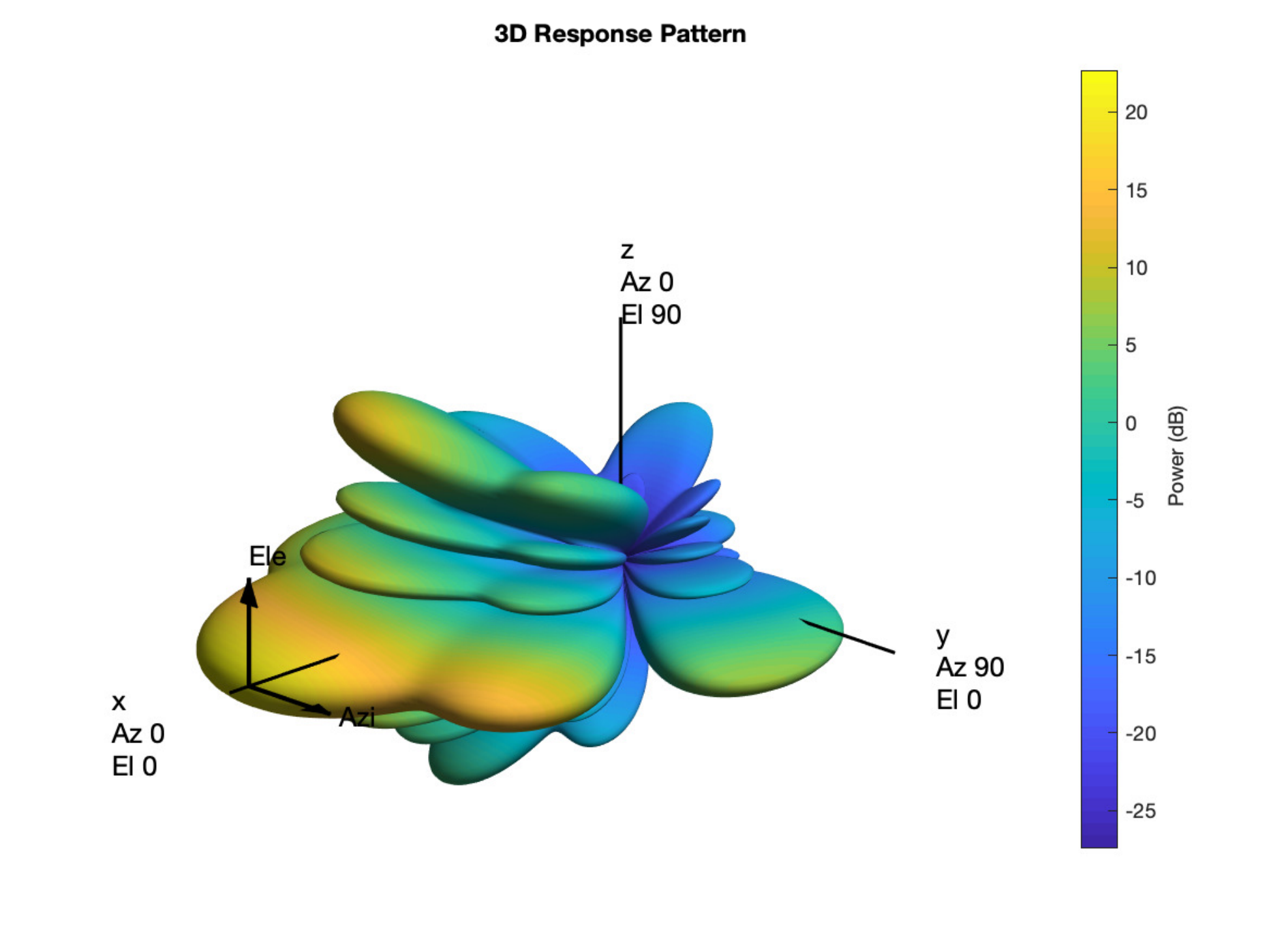}
			\caption{Beam Pattern}
			\label{pattern}
		\end{subfigure}%
		\begin{subfigure}{.5\textwidth}
			\centering
			\includegraphics[width=1.0\linewidth]{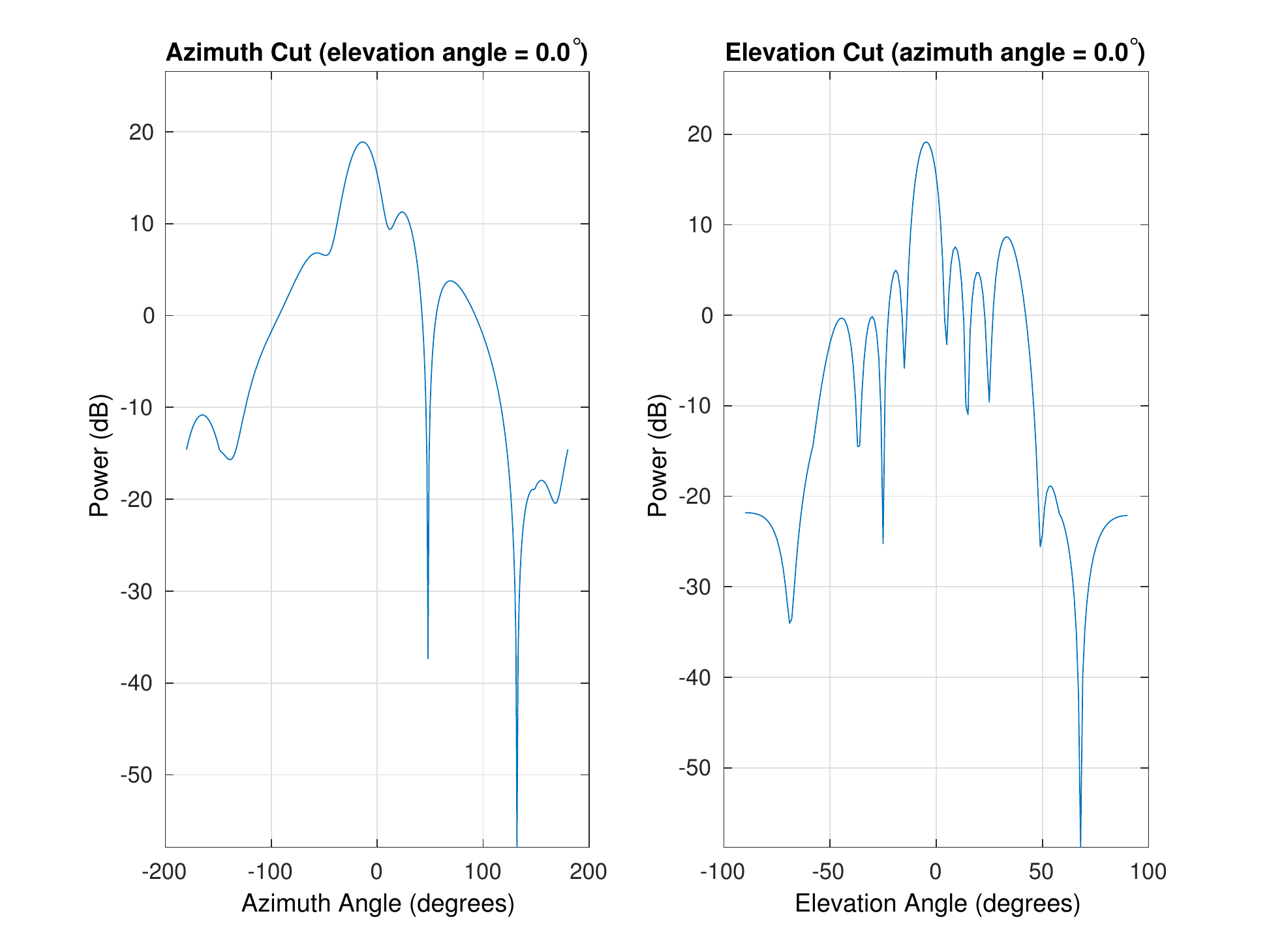}
			\caption{Elevation and Azimuth Cuts}
			\label{Scenario2}
		\end{subfigure}
		\caption{Beam pattern corresponding to a typical RL action.}
		\label{Patterns_and_actions}
	\end{figure}

\begin{figure}
		\centering
		\begin{subfigure}{.5\textwidth}
			\centering
    		\includegraphics[width=1.0\linewidth]{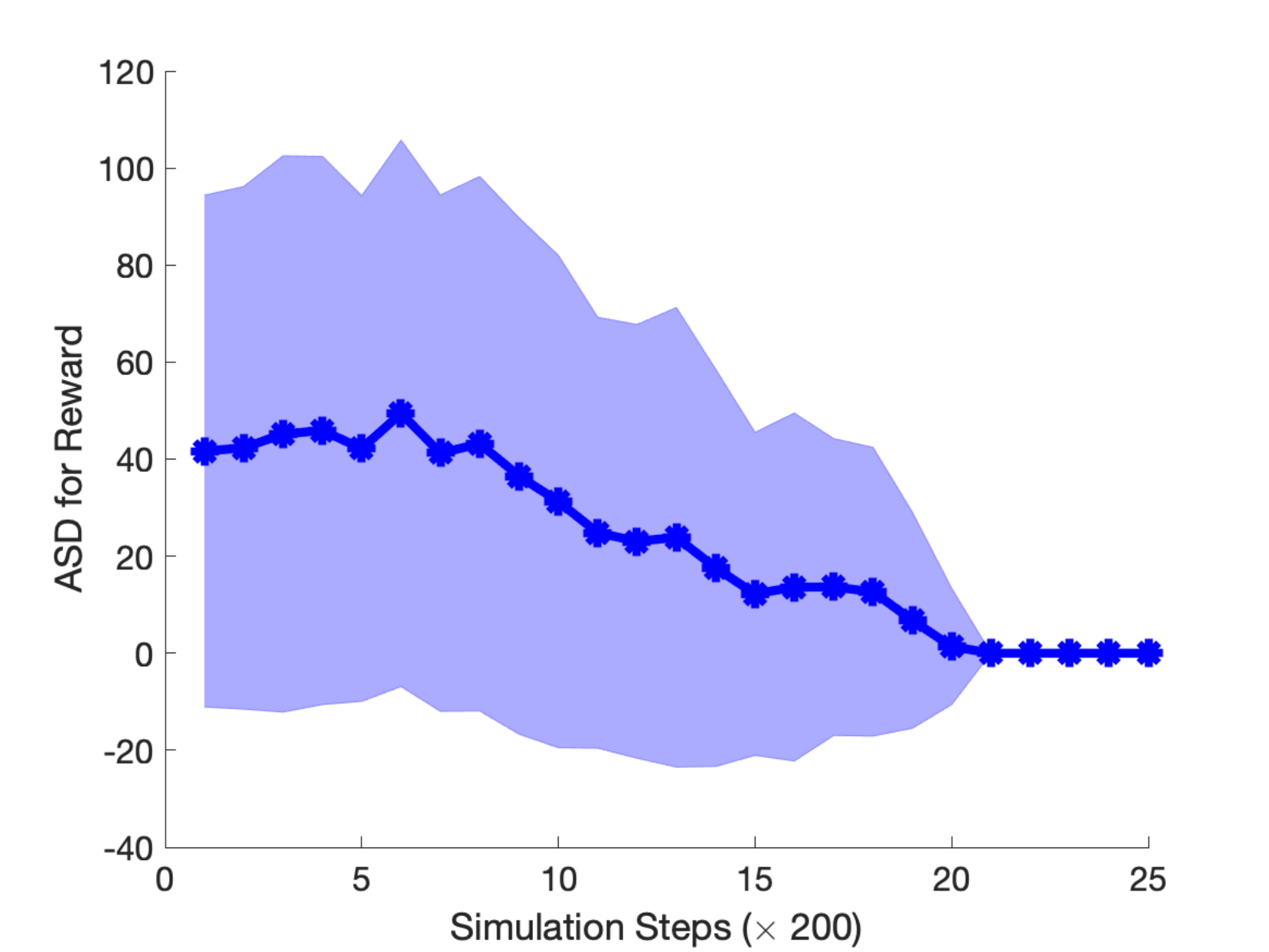}
    		\caption[width=0.5\linewidth]{{ASD in reward from DRL agent and Oracle.}}
    		%\captionsetup{justification=centering}
    		\label{1Cell_Reward}
		\end{subfigure}%
		\begin{subfigure}{.5\textwidth}
			\centering
    		\includegraphics[width=1.0\linewidth]{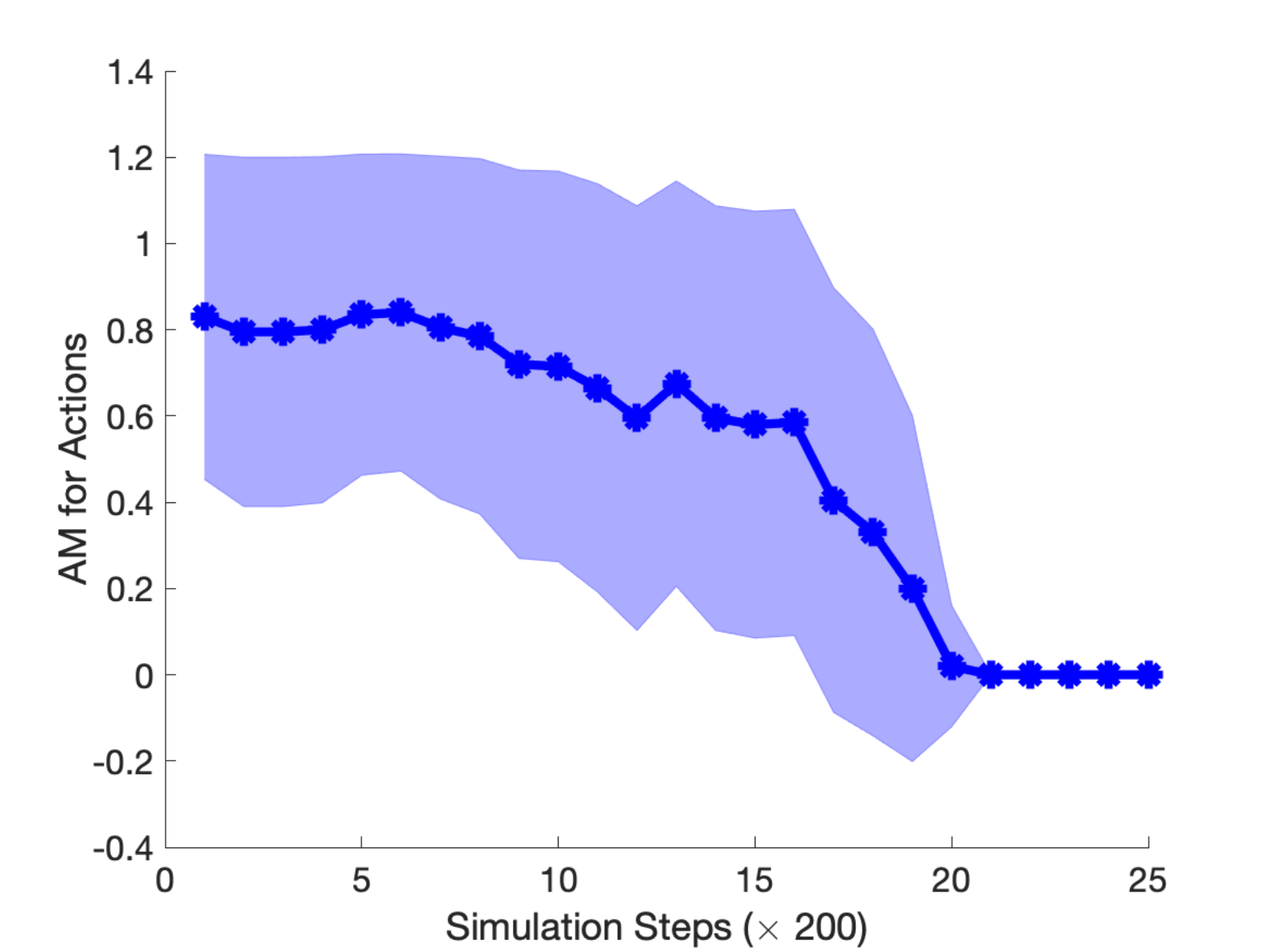}
    		%\captionsetup{margin= {20pt},justification=centerlast,skip=-5pt,font=normalsize}
    		%\vspace{0.1cm}
    		\caption[width=0.5\linewidth]{{Average action mismatch with Oracle. }}
    		%\captionsetup{justification=centering}
    		\label{1Cell_BS1}
		\end{subfigure}
		\caption{Results for periodic mobility pattern in a single sector dynamic environment: (a) average squared difference (ASD) between reward  achieved by DRL agent and the reward obtained by Oracle; (b) average mismatch (AM) between actions taken by the DRL agents and the Oracle.}
		\label{singlecell_results}
	\end{figure}

	At each time step, the RL agent has 5 actions to choose from, i.e., there are 5 different beam weight vectors available for the agent. Each of the actions corresponds to a unique beam pattern. As an illustration, one such beam pattern and the associated elevation and azimuth cuts are shown in Fig.~\ref{Patterns_and_actions}.
	Based on the change in user distribution, the agent adaptively selects the beam that maximizes the total number of connected UEs. Figure \ref{1Cell_Reward} shows the average squared difference (ASD) between the reward (total number of connected UEs) obtained by the DRL agent and the reward predicted by Oracle:
	\begin{align}
	\text{ASD}=\frac{1}{N'}\sum_{n=1}^{N'}\left(R_n^{\text{Agent}}-R_n^{\text{Oracle}}\right)^2,
	\end{align}
	where $R_n^{\text{Agent}}$ and $R_n^{\text{Oracle}}$ denote instantaneous reward at $n$-th time step obtained by DRL agent and Oracle, respectively; $N'$ represent the number of time steps used for averaging. \textit{Oracle} is defined as an entity which has the complete and perfect knowledge of the environment and user distribution; it is essentially an exhaustive search method in order to compute the maximum attainable reward at any given scenario. Each point in Fig.~\ref{1Cell_Reward}  represent ASD over $N'= 200$ time steps. In Fig.~\ref{1Cell_Reward}, we have also shown the shaded error bar, which represent the maximum difference from the mean value within every $N'$ time steps. It can be observed that at the beginning of training, ASD between rewards obtained by the RL agent and the Oracle is quite high. However, as time goes by, ASD gradually decreases, and finally, at the completion of training, rewards from RL agent converges completely with that from Oracle. This is due to the fact that at the beginning of training, the agent explores different actions and collects the memory. During the exploration phase, the agent tries out all available actions, and attempts to learn the optimal beam weights for different user distributions.  Over time, this exploration rate decreases, and exploitation increases, i.e.,  agent tends to choose more frequently the best known actions so far that maximize the reward.
	
% 	\begin{figure}[h!]
% 		\centering
% 		\includegraphics[width=0.7\linewidth]{Figures/1Cell/singleCell_Reward}
% 		%\captionsetup{margin= {20pt},justification=centerlast,skip=-5pt,font=normalsize}
% 		%\vspace{0.1cm}
% 		\caption[width=0.5\linewidth]{{Average squared difference in reward from DRL agent and Oracle.}}
% 		%\captionsetup{justification=centering}
% 		\label{1Cell_Reward}
% 	\end{figure}
	
	Fig.~\ref{1Cell_BS1} shows the results for average mismatch (AM) in actions (selected beam pattern) taken by the DRL agent and the Oracle, respectively, where AM is defined as  
\begin{align}
\text{AM}= \frac{1}{N'}\sum_{n=1}^{N'}\mathds{1}_{\left(A_n^{\text{Agent}} \neq A_n^{\text{Oracle}}\right)},
\end{align}
% 	\begin{figure}[h!]
% 		\centering
% 		\includegraphics[width=0.7\linewidth]{Figures/1Cell/SingleCell_BS1_actions}
% 		%\captionsetup{margin= {20pt},justification=centerlast,skip=-5pt,font=normalsize}
% 		%\vspace{0.1cm}
% 		\caption[width=0.5\linewidth]{{Average action mismatch with Oracle. }}
% 		%\captionsetup{justification=centering}
% 		\label{1Cell_BS1}
% 	\end{figure}
where $A_n^{\text{Agent}}$ and $A_n^{\text{Oracle}}$ denote the actions selected for $n$-th time step by the DRL agent and the Oracle, respectively, and the indicator function,  $\mathds{1}_{\left(A_n^{\text{Agent}} \neq A_n^{\text{Oracle}}\right)}$, is defined as
\begin{align}
\mathds{1}_{\left(A_n^{\text{Agent}} \neq A_n^{\text{Oracle}}\right)}= \begin{cases}
1, & \text{if  } A_n^{\text{Agent}} \neq A_n^{\text{Oracle}}\\
0,  & \text{if  } A_n^{\text{Agent}} = A_n^{\text{Oracle}}.
\end{cases}
\end{align}
%\begin{figure}[h!]
%	\centering
%	\includegraphics[width=0.8\linewidth]{Figures/1Cell/1Cell_BS1_zoomed}
%	%\captionsetup{margin= {20pt},justification=centerlast,skip=-5pt,font=normalsize}
%	%\vspace{0.1cm}
%	\caption[width=0.5\linewidth]{{Action at convergence for singles sector case}}
%	%\captionsetup{justification=centering}
%	\label{1Cell_BS1_zoomed}
%\end{figure}
It can be observed that action mismatch is quite large at the start of the training because of high exploration rate. However, at the end of training phase, actions taken by the DRL agent and the Oracle converge completely, and average mismatch reduces to zero.

	\subsection{ Results for multiple sector dynamic environment:}
\begin{figure}
		\centering
		\begin{subfigure}{.5\textwidth}
			\centering
			\includegraphics[width=1.0\linewidth]{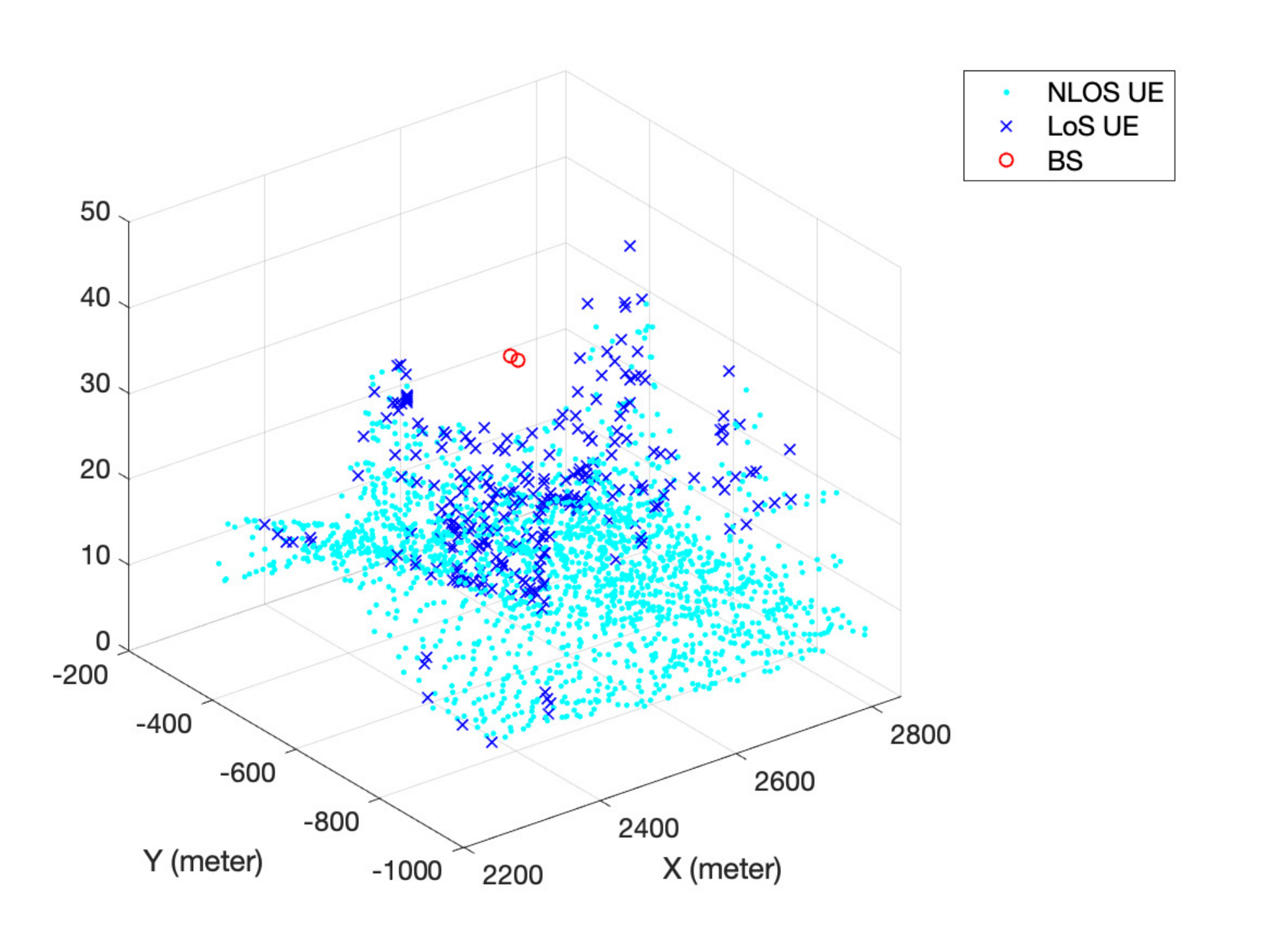}
			\caption{Scenario 1}
			\label{Scenario1}
		\end{subfigure}%
		\begin{subfigure}{.5\textwidth}
			\centering
			\includegraphics[width=1.0\linewidth]{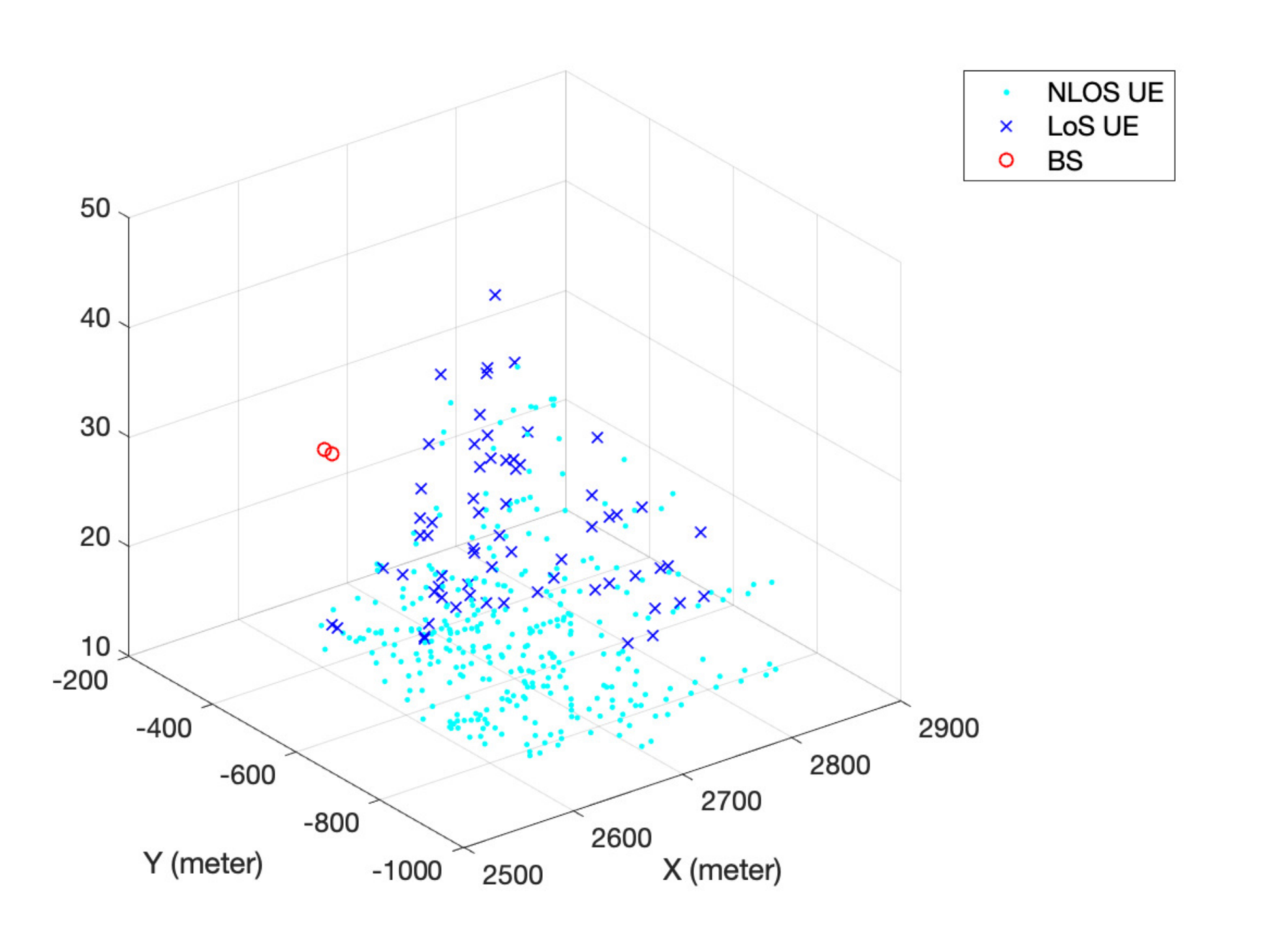}
			\caption{Scenario 2}
			\label{Scenario2}
		\end{subfigure}
		\begin{subfigure}{.5\textwidth}
			\centering
			\includegraphics[width=1.0\linewidth]{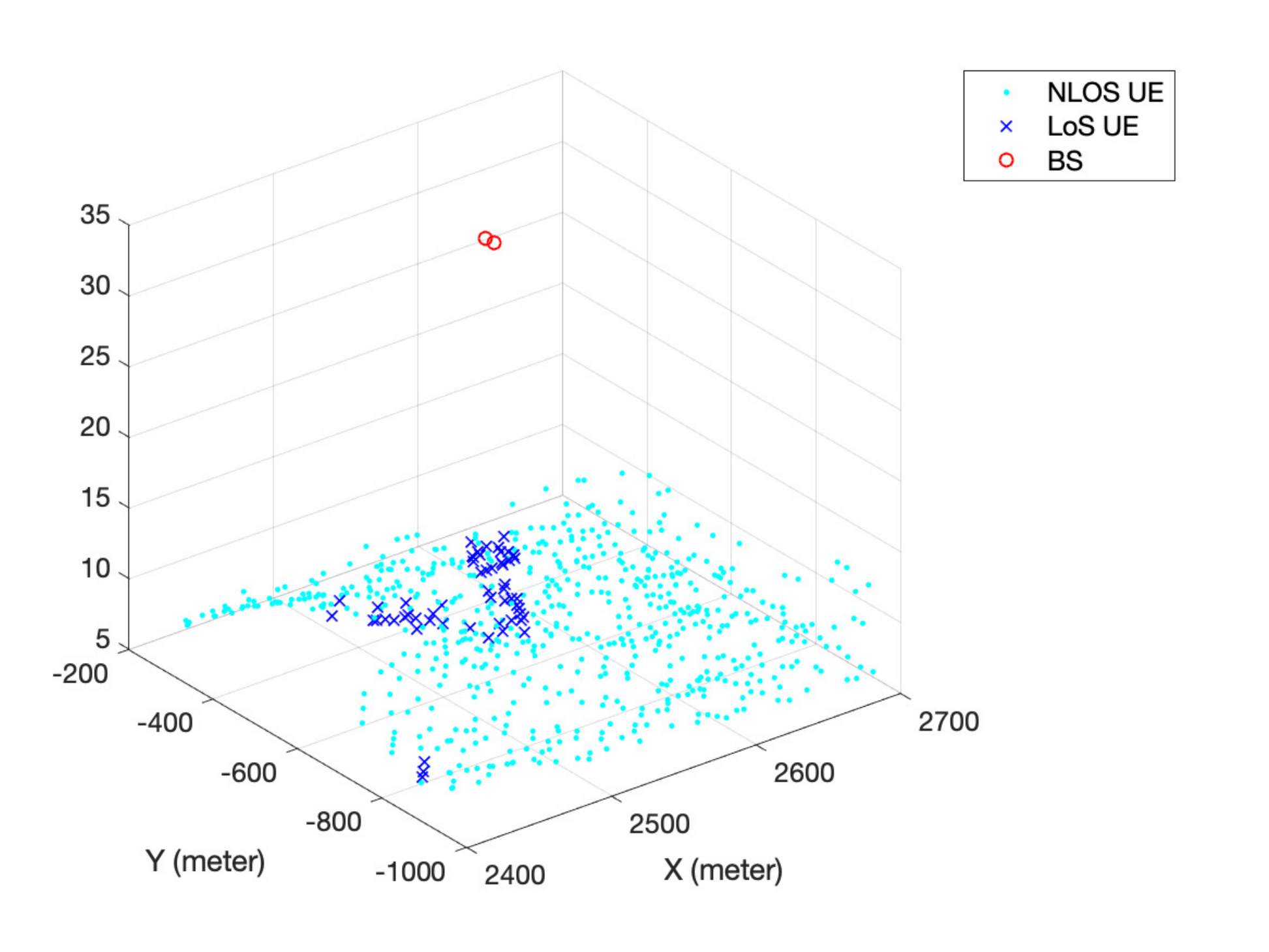}
			\caption{Scenario 3}
			\label{Scenario3}
		\end{subfigure}
		\caption{Users' Distribution Patterns for 3 Scenarios.}
		\label{Scenarios}
	\end{figure}

\begin{figure}
		\centering
		\begin{subfigure}{.5\textwidth}
			\centering
    		\includegraphics[width=1.0\linewidth]{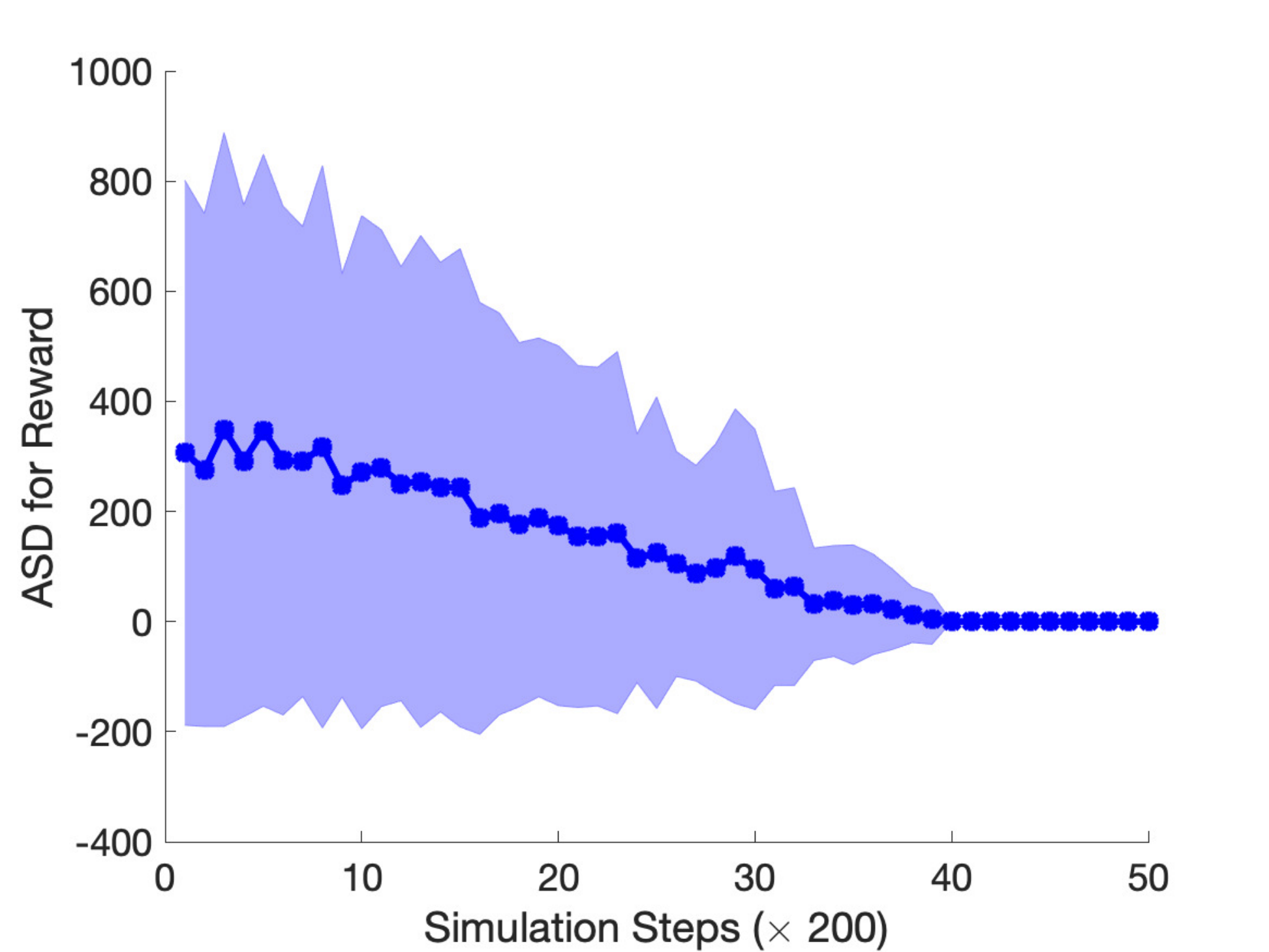}
    		%\captionsetup{margin= {20pt},justification=centerlast,skip=-5pt,font=normalsize}
    		%\vspace{0.1cm}
    		\caption[width=0.5\linewidth]{{}}
    		%\captionsetup{justification=centering}
    		\label{3scenarios}
		\end{subfigure}%
		\begin{subfigure}{.5\textwidth}
			\centering
    		\includegraphics[width=1.0\linewidth]{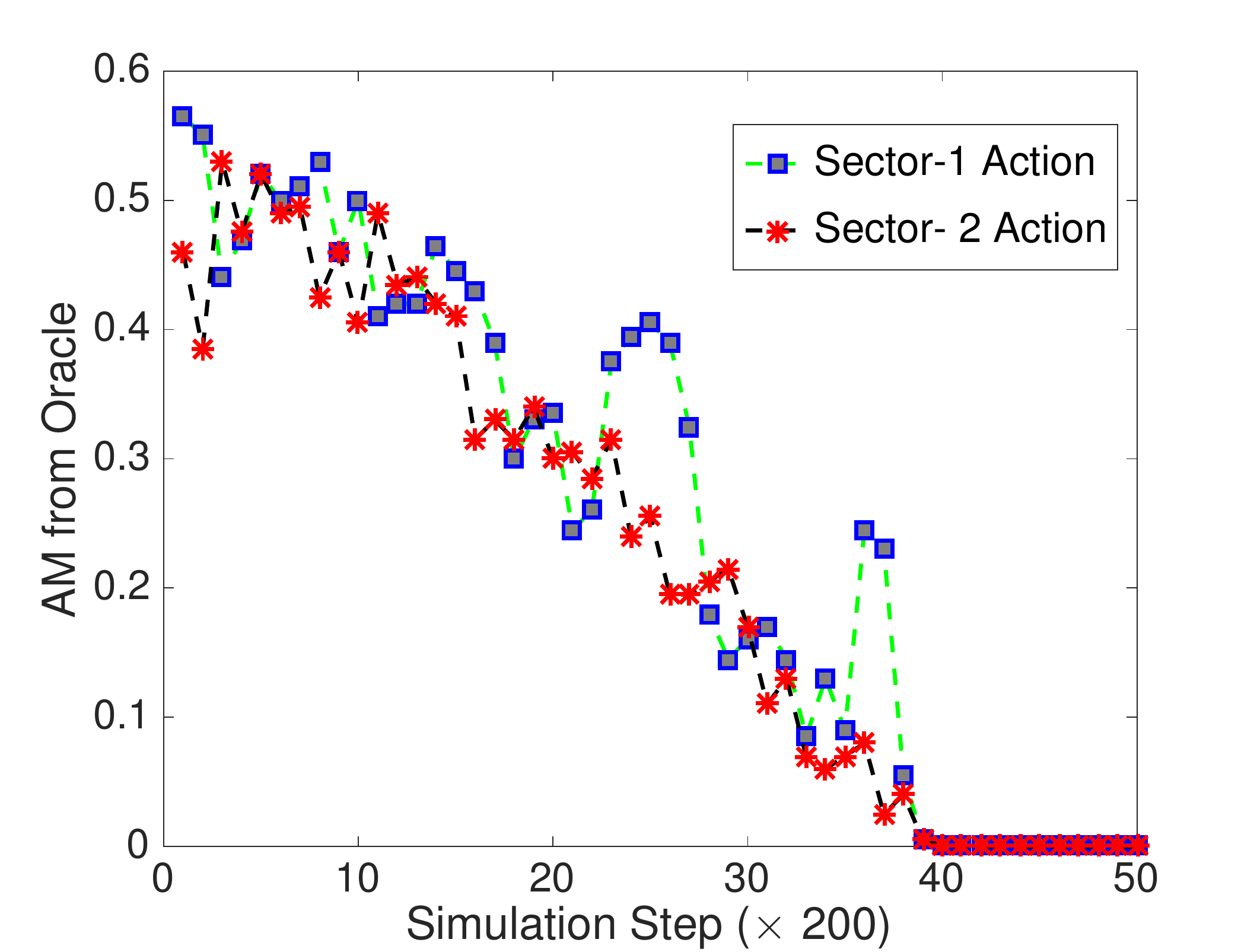}
    		%\captionsetup{margin= {20pt},justification=centerlast,skip=-5pt,font=normalsize}
    		%\vspace{0.1cm}
    		\caption[width=0.5\linewidth]{{}}
    		%\captionsetup{justification=centering}
    		\label{Multiple_Action}
		\end{subfigure}
		\caption[width=0.5\linewidth]{Results for periodic mobility pattern in a multiple sector dynamic environment: (a) average squared difference (ASD) between reward  achieved by DRL agent and the reward obtained by Oracle; (b) average mismatch (AM) between actions taken by DRL agents for each sector and the corresponding Oracles.}
		\label{multipleecell_results}
	\end{figure}

\begin{figure}
		\centering
		\begin{subfigure}{.5\textwidth}
			\centering
			\includegraphics[width=1.0\linewidth]{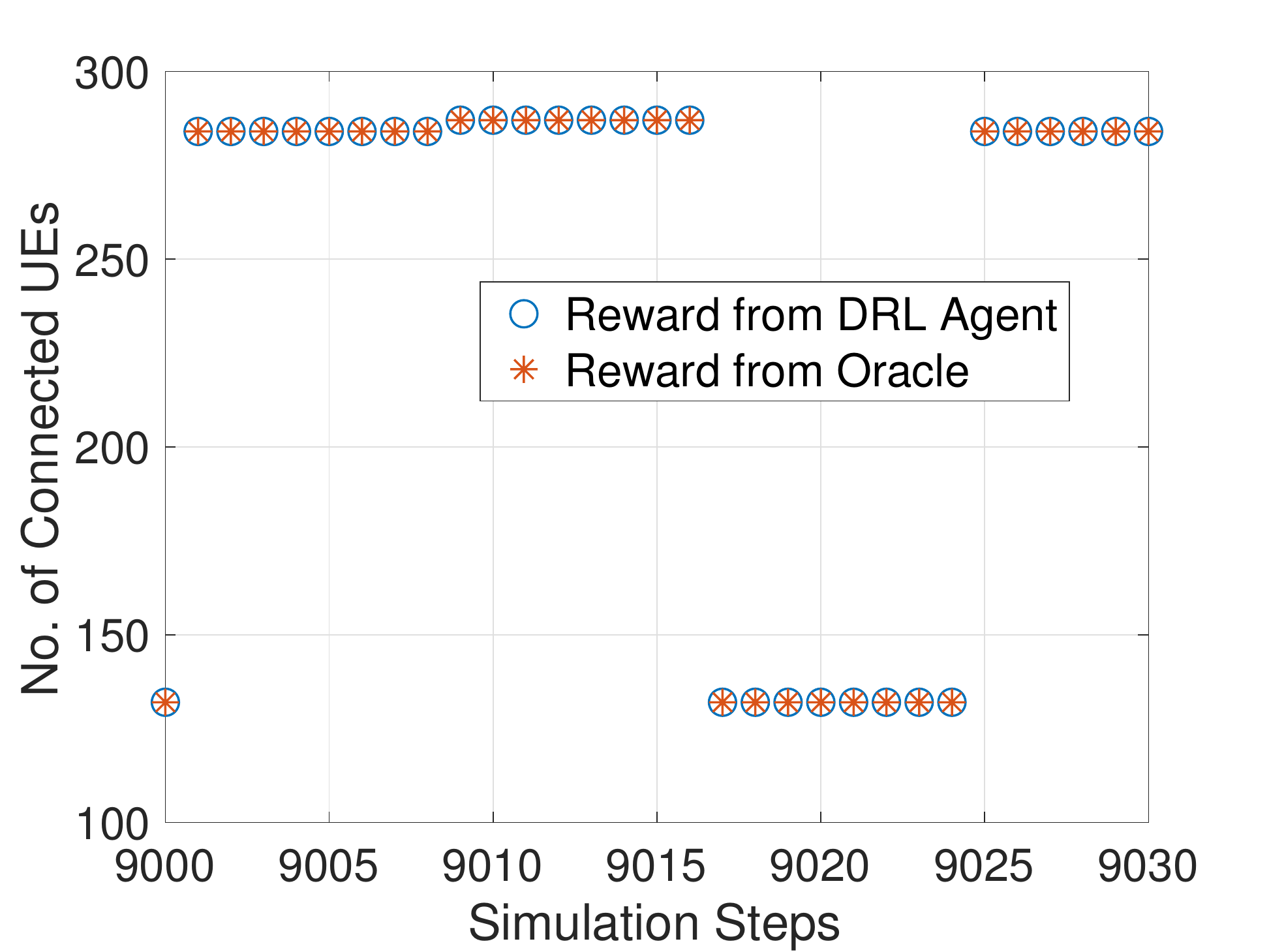}
			\caption{Rewards}
			\label{3scen_reward_zoomed}
		\end{subfigure}%
		\begin{subfigure}{.5\textwidth}
			\centering
			\includegraphics[width=1.0\linewidth]{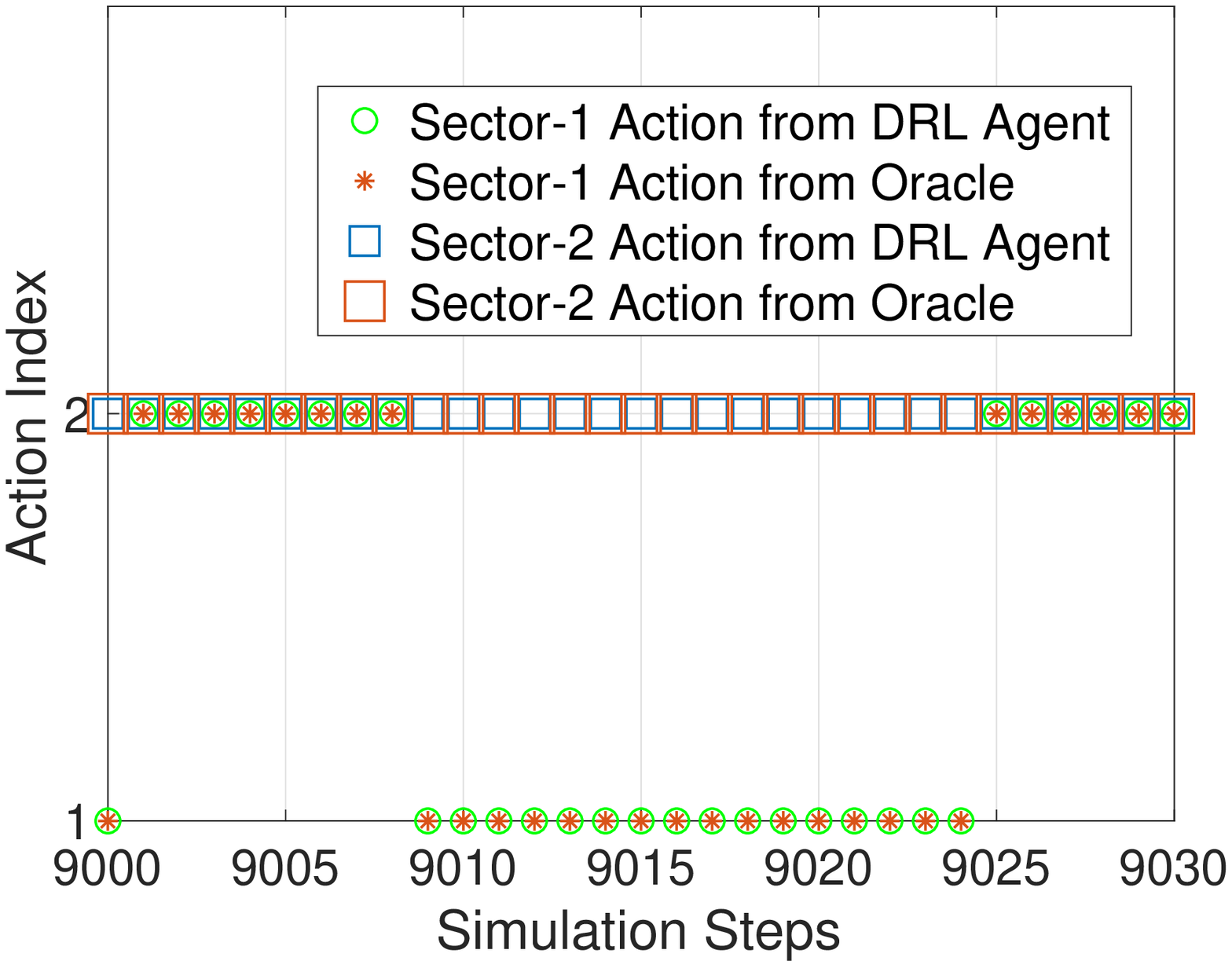}
			\caption{Actions}
			\label{3scen_BS1_zoomed}
		\end{subfigure}
		\caption{Instantaneous reward (a) and instanteneous actions (b) at Convergence for multiple sectors environment and periodic user mobility pattern.}
		\label{3scen_Instantaneous}
	\end{figure}

	In this sub-section, we present the simulation results for multiple sector dynamic environment.
	We consider two sectors, each at a height of $35$ m from ground. Each sector has two possible beam patterns to choose from. Three scenarios are considered this time based on users' location coordinate:  \textbf{Scenario-1}: $X \geq 2600 \text{ m}, Z\ge10 \text{ m}$; \textbf{Scenario-2}: $2400 \leq X \leq 2700 \text{ m}, Z\le 12 \text{ m}$; \textbf{Scenario-3}: $X \leq 2500 \text{ m}, Z \leq 15 \text{ m}$. The scenarios with  line of sight (LoS) and non-line of sight (NLoS) UEs are shown in Fig.~\ref{Scenarios}. 
	We assume the scenarios periodically change every 8 time steps. The agent is responsible for simultaneously selecting the optimal beam patterns for both sectors for maximizing the number of connected UEs in the network. The average squared difference in rewards achieved by the agent and the oracle for multiple sectors scenario is shown in Fig.~\ref{3scenarios}. Similarly to single cell case, as training increases, overall rewards attained by the agent and the oracle converge  completely. In other words, the agent is able to dynamically optimize the beam patterns for both sectors simultaneously in the interference environment, and maximize the overall rewards from the network in all scenarios or user distributions. In Fig.~\ref{Multiple_Action}, we show the average action mismatch for both sectors. It can be observed that towards the end of exploration phase, average action mismatches between the sectors and the corresponding Oracles reduce to zero. The instantaneous rewards and actions at convergence of the algorithm are shown in Fig.~\ref{3scen_Instantaneous}, where, for clarity, we zoom in for time steps between 9000 and 9030. We can observe that scenarios change every 8 time steps and maximum number of connected UEs are different for the three scenarios. Optimal strategy for sector-1 is to select action 2 while in scenario 1, and select action 1 while in scenario 2 or 3. On the other hand, optimal strategy for sector-2 is to select action 2 for all scenarios. In reinforcement learning, it is, in general, difficult to obtain convergence if the reward values are too close. However, we can observe from Fig.~\ref{3scen_reward_zoomed} that the DRL agent can completely converge with the oracle and take the corresponding best actions even when the reward values for scenario 1 and scenario 2 differ by only a small number. This indicates the accuracy of self-tuning sectorization strategy developed in this work.
	
\subsection{Multiple sectors environment with Markovian mobility pattern}

In this sub-section, we present the performance analysis for DRL-based self-tuning beamforming in multiple sector environment and for the case where user distributions alternate between two scenarios following a Markovian mobility pattern. It is to be noted here that, in general, users' mobility pattern has some intrinsic regularity. For example, users can be clustered more in the commercial area during day time while they move to residential are in the evening. Hence, the periodic mobility patterns considered in the previous two sub-sections rather closely depict the actual mobility pattern in cellular network. Nevertheless, in this sub-section, we consider the Markovian mobility in order to verify the robustness of the developed self-tuning sectorization algorithm for the extreme case when users' mobility pattern doesn't have any regularity and users move between different scenarios in a random fashion.

\begin{figure}[h!]
	\centering
	\includegraphics[width=0.7\linewidth]{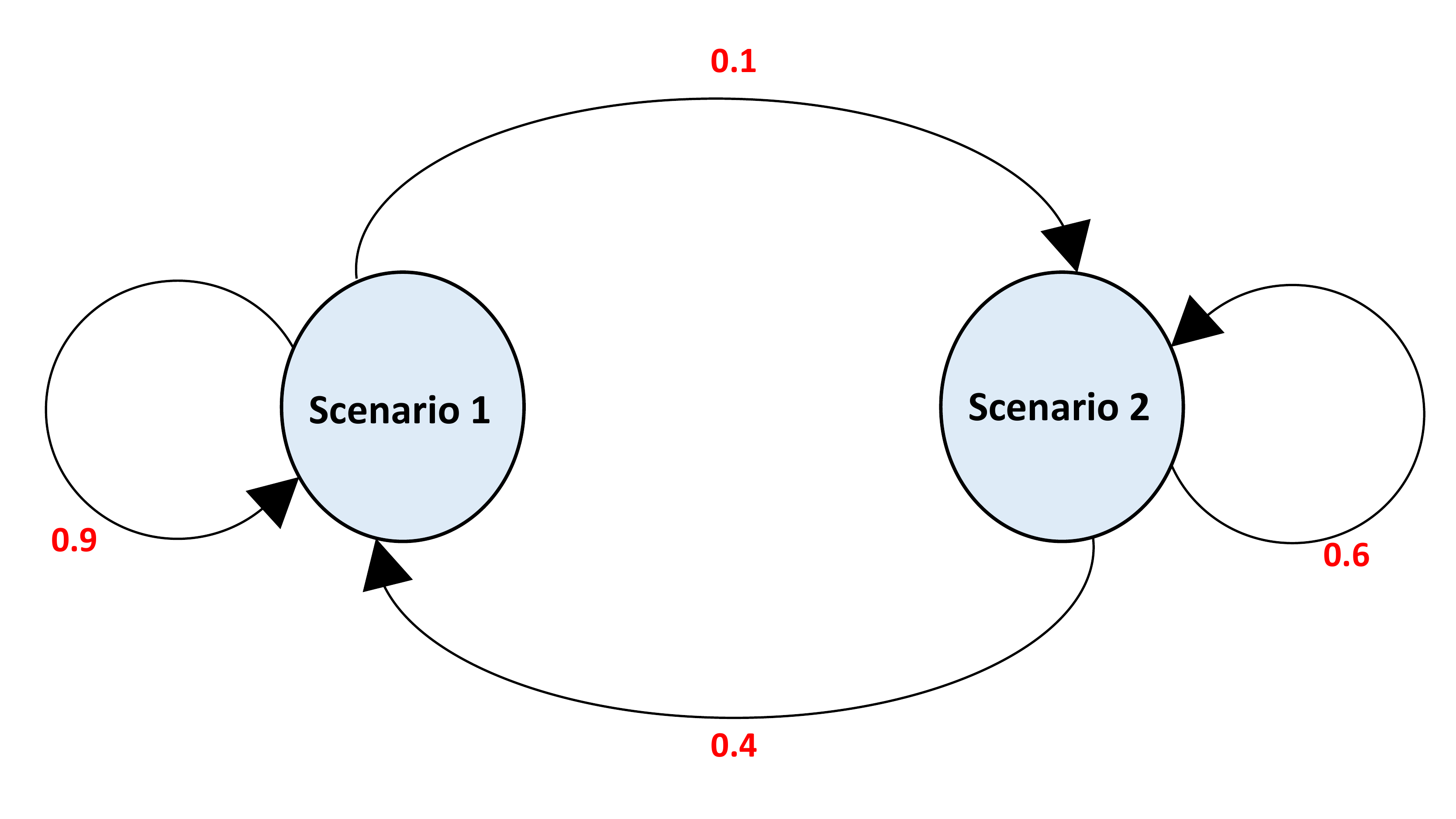}
	%\captionsetup{margin= {20pt},justification=centerlast,skip=-5pt,font=normalsize}
	%\vspace{0.1cm}
	\caption[width=0.5\linewidth]{{State Transition Diagram for Markov Mobility Pattern }}
	%\captionsetup{justification=centering}
	\label{transition}
\end{figure}

\begin{figure}
		\centering
		\begin{subfigure}{.5\textwidth}
			\centering
        	\includegraphics[width=1.0\linewidth]{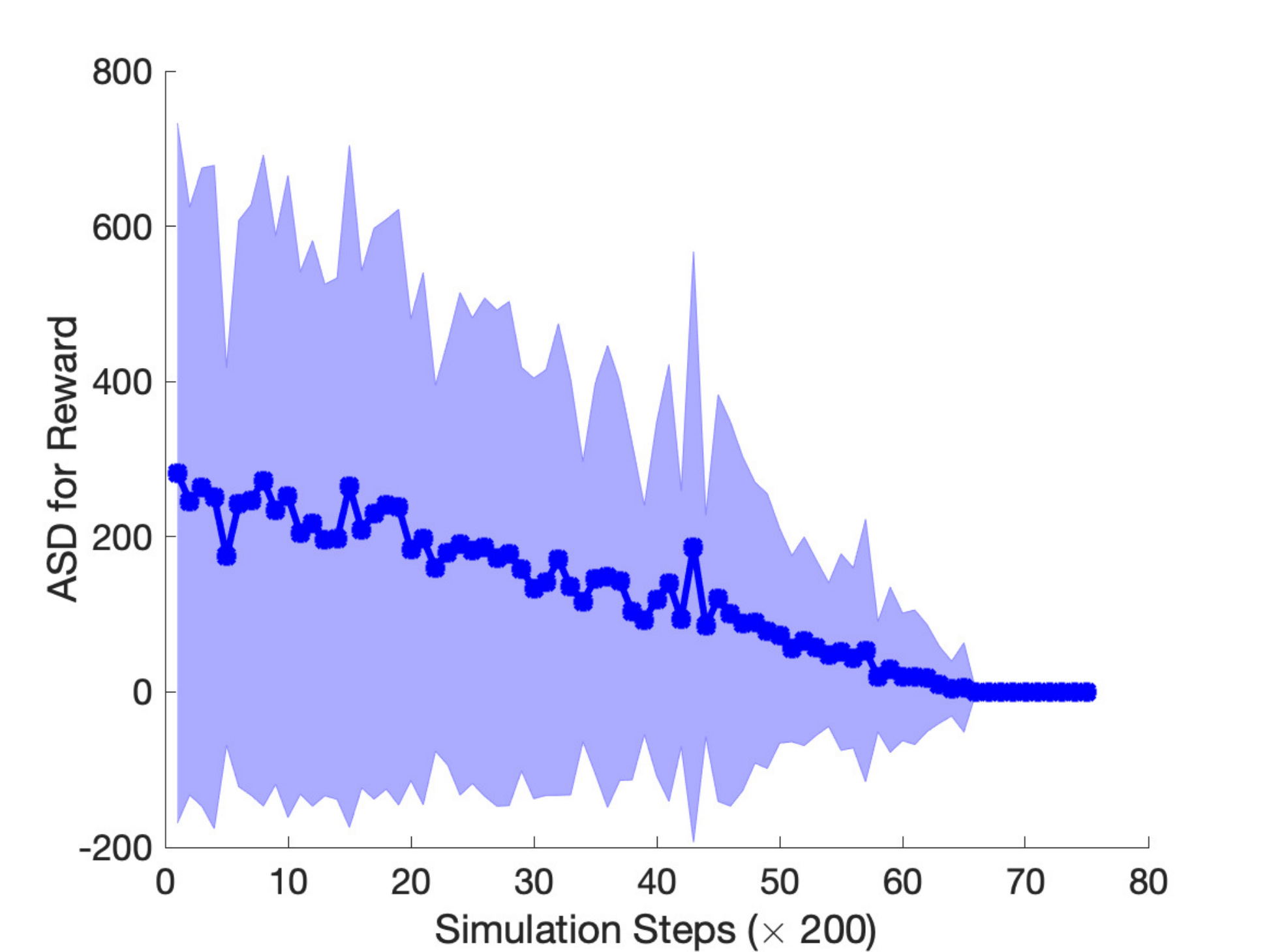}
        	%\captionsetup{margin= {20pt},justification=centerlast,skip=-5pt,font=normalsize}
        	%\vspace{0.1cm}
        	\caption[width=0.5\linewidth]{{}}
        	%\captionsetup{justification=centering}
        	\label{markov_reward}
		\end{subfigure}%
		\begin{subfigure}{.5\textwidth}
			\centering
        	\includegraphics[width=1.0\linewidth]{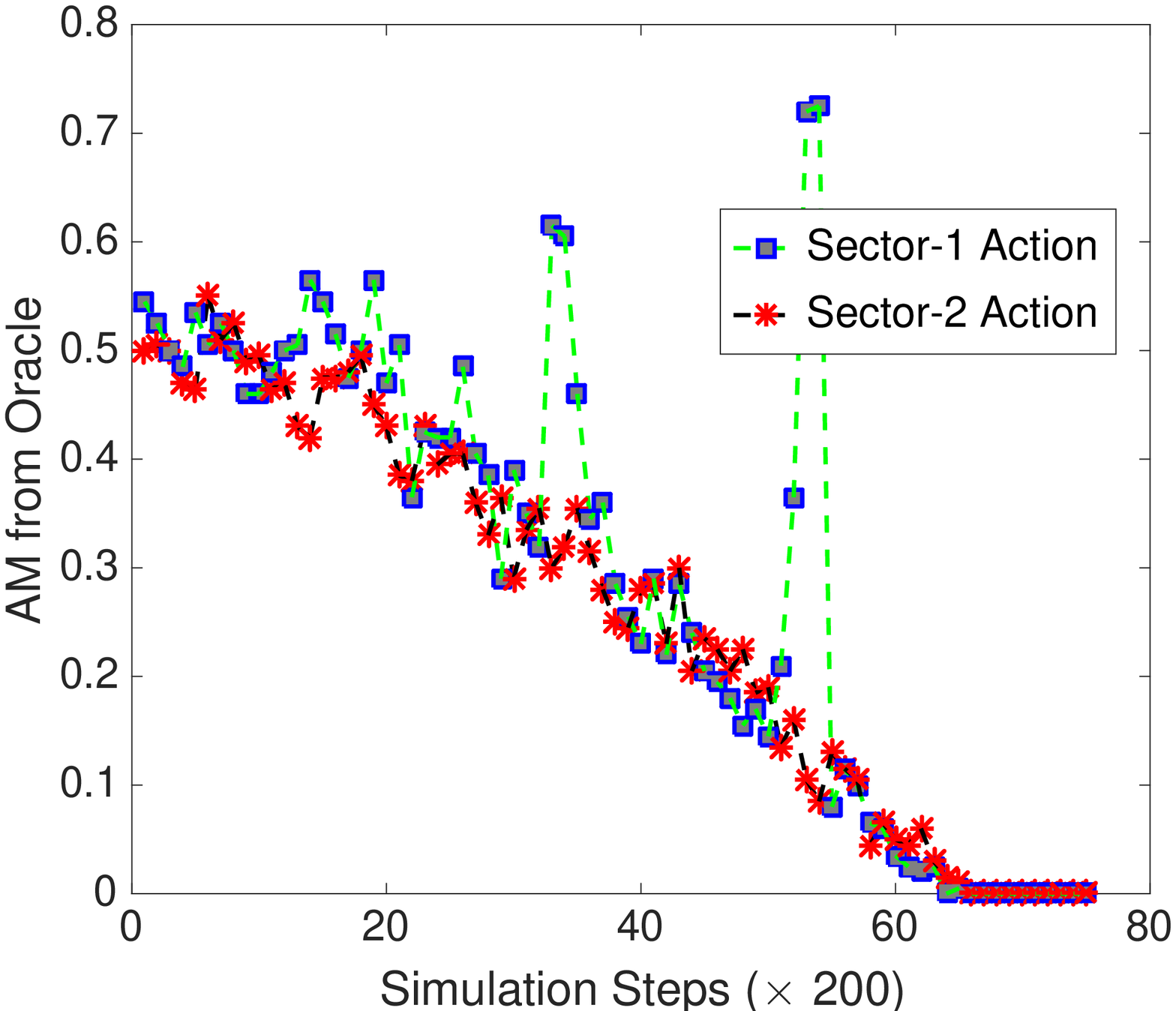}
        	%\captionsetup{margin= {20pt},justification=centerlast,skip=-5pt,font=normalsize}
        	%\vspace{0.1cm}
        	\caption[width=0.5\linewidth]{{}}
        	%\captionsetup{justification=centering}
        	\label{markov_action}
		\end{subfigure}
		\caption[width=0.5\linewidth]{Results for Markov mobility pattern in a multiple sector dynamic environment: (a) average squared difference (ASD) between reward  achieved by DRL agent and the reward obtained by Oracle; (b) average mismatch (AM) between actions taken by DRL agent for each sector and the corresponding Oracles.}
		\label{multipleecell_results}
	\end{figure}

\begin{figure}
	\centering
	\begin{subfigure}{.5\textwidth}
		\centering
		\includegraphics[width=1.0\linewidth]{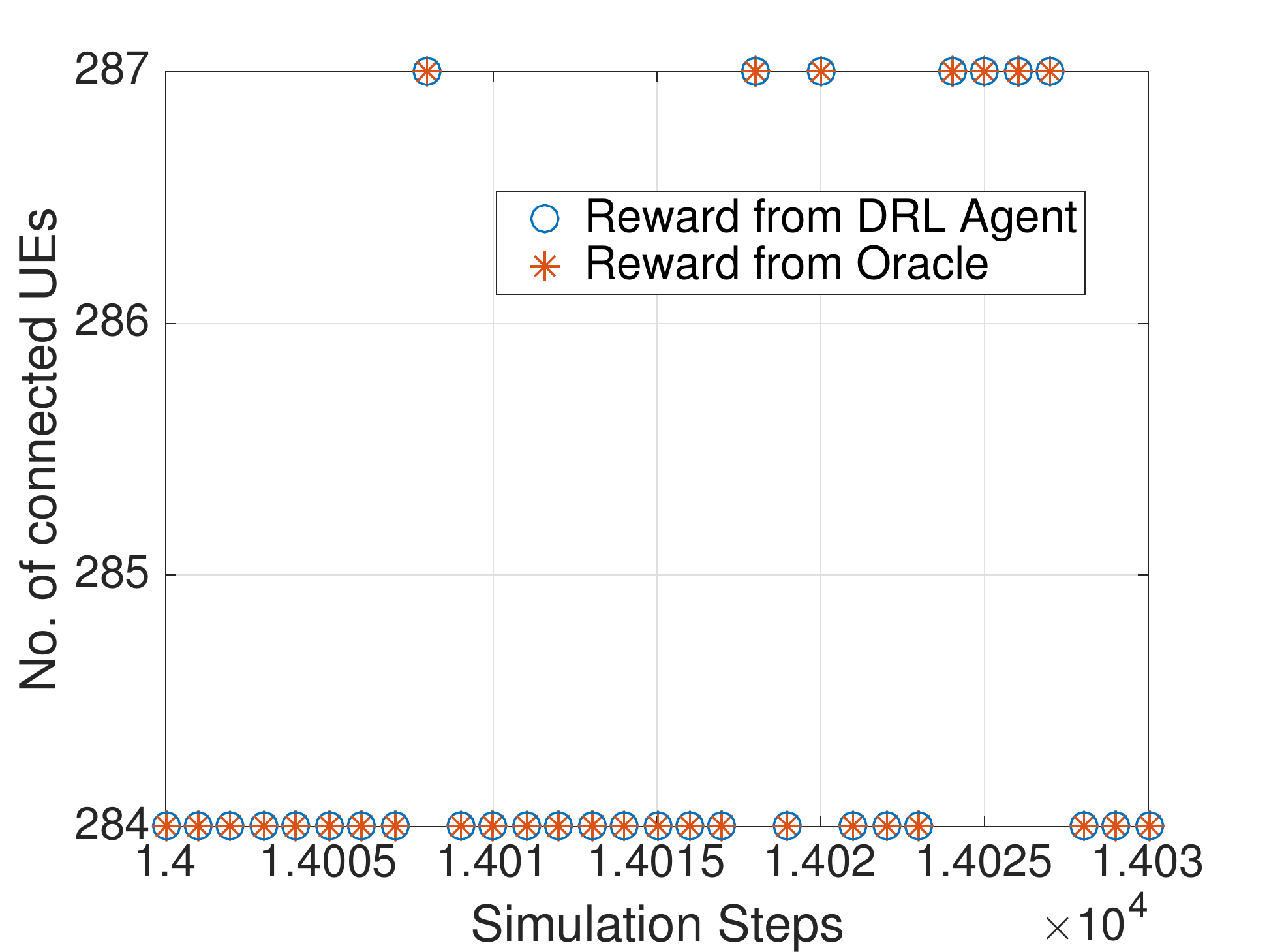}
		\caption{Rewards}
		\label{Markov_reward_zoomed}
	\end{subfigure}%
	\begin{subfigure}{.5\textwidth}
		\centering
		\includegraphics[width=1.0\linewidth]{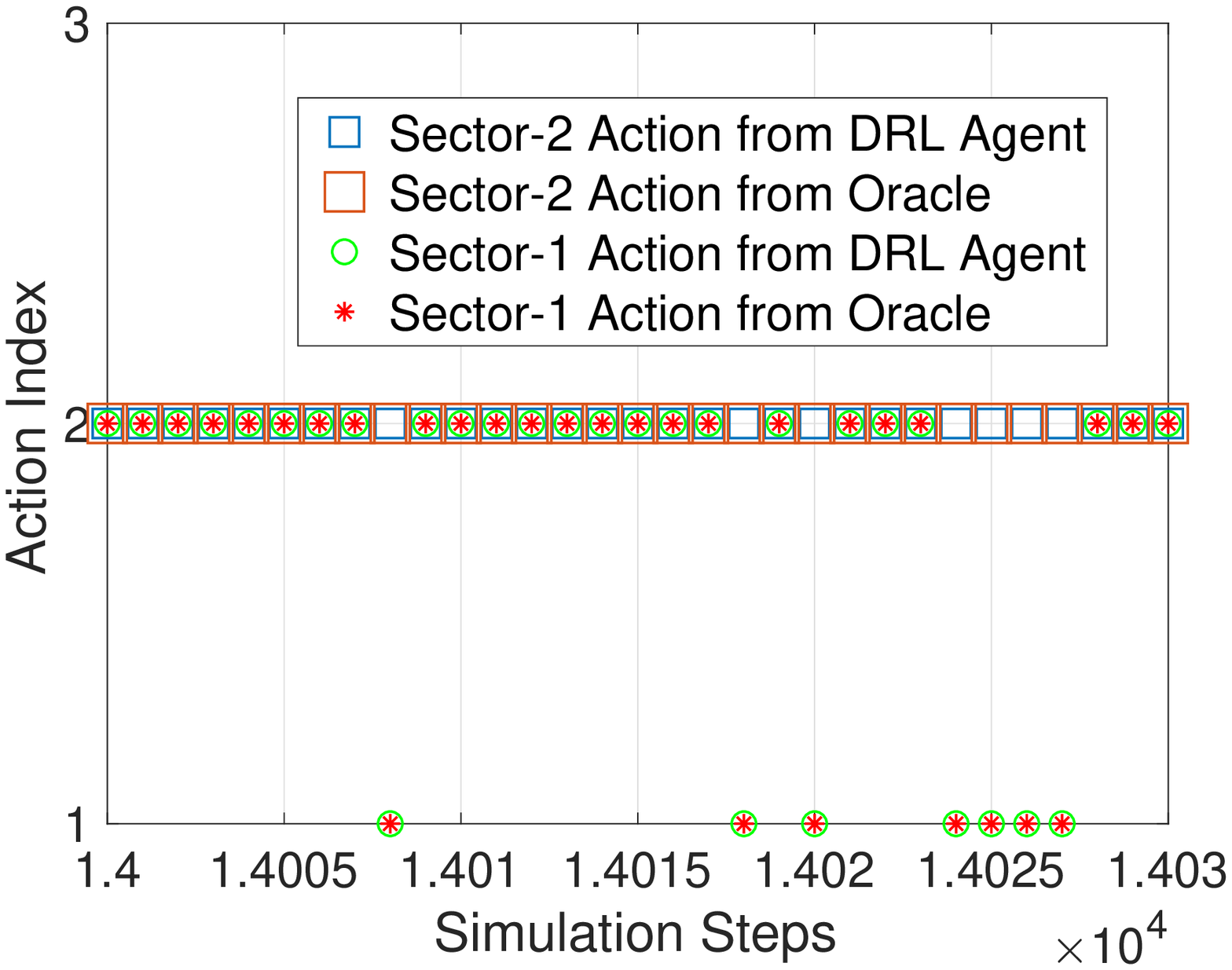}
		\caption{Actions}
		\label{Markov_BS1_zoomed}
	\end{subfigure}
	\caption{Instantaneous reward (a) and instanteneous actions (b) at Convergence for multiple sectors environment and Markov user mobility pattern.}
	\label{Markov_Instantaneous}
\end{figure}

We consider two scenarios defined similarly to the ones in Section \ref{simulation_singlecell}, and assume the users' locations switch between these two scenarios with transition probability governed by the state transition diagram shown in Fig.~\ref{transition}.
Moreover, we consider two sectors each having two possible beam patterns to choose from for each scenario. Fig.~\ref{markov_reward} shows the average squared difference for rewards attained by the RL agent and the oracle for Markov mobility pattern. We can observe that similarly to the periodic cases presented in previous two subsections,  RL agent does converge with the oracle even for probabilistic mobility, and ASD goes to zero after the training phase. Average mismatch in actions between the sectors and the corresponding oracles are shown in Fig.~\ref{markov_action}. It can be seen that average mismatch in actions for both sectors reduce to zero at the end of the training phase. Finally, the  instantaneous rewards  achieved and the actions taken by the sectors at convergence of the algorithm are shown in Fig.~\ref{Markov_Instantaneous}, which, again, indicates perfect convergence for Markov mobility pattern in multiple cell environment.

% \begin{figure}[h!]
% 	\centering
% 	\includegraphics[width=0.7\linewidth]{Figures/Markov/markov_reward}
% 	%\captionsetup{margin= {20pt},justification=centerlast,skip=-5pt,font=normalsize}
% 	%\vspace{0.1cm}
% 	\caption[width=0.5\linewidth]{{Average Squared Error Rewards for Markov Environment }}
% 	%\captionsetup{justification=centering}
% 	\label{markov_reward}
% \end{figure}

%\begin{figure}[h!]
%	\centering
%	\includegraphics[width=0.8\linewidth]{Figures/Markov/reward_zoomed}
%	%\captionsetup{margin= {20pt},justification=centerlast,skip=-5pt,font=normalsize}
%	%\vspace{0.1cm}
%	\caption[width=0.5\linewidth]{{Rewards for Markov Environment at Convergence}}
%	%\captionsetup{justification=centering}
%	\label{Markov_reward_zoomed}
%\end{figure}

% \begin{figure}[h!]
% 	\centering
% 	\includegraphics[width=0.7\linewidth]{Figures/Markov/markov_action}
% 	%\captionsetup{margin= {20pt},justification=centerlast,skip=-5pt,font=normalsize}
% 	%\vspace{0.1cm}
% 	\caption[width=0.5\linewidth]{{Base Stations' Action Mismatch for Markov Environment }}
% 	%\captionsetup{justification=centering}
% 	\label{markov_action}
% \end{figure}

	\section{Conclusion}\label{conclusion}
	In this work, we have developed a framework for self-tuning cell sectorization through MIMO broadcast beam optimization using  deep reinforcement learning. To be specific, we have proposed learning strategies for both single sector and multiple sectors environment with dynamic user distribution.
	The proposed solutions can autonomously and adaptively update the RF parameters based on the changes in user distributions. Simulation results show that the proposed DRL-based method completely converges with the Oracle-suggested optimal solutions for both periodic and Markovian user mobility patterns.
	
%\bibliographystyle{IEEEtran}
%\bibliography{IEEEabrv,MIMO_OFDM}

\end{document}